\documentclass{article}

\usepackage{amssymb}

\usepackage{graphicx}

\begin{document}

\title{Studies in a Random Noise Model of Decoherence } 

\author{P. Korcyl, J. Wosiek \\
M. Smoluchowski Institute of Physics\\
 Jagellonian
University, Reymonta 4, 30-059 Cracow, Poland\\
\\
L.~Stodolsky\\
Max-Planck-Institut f\"ur Physik
(Werner-Heisenberg-Institut)\\
F\"ohringer Ring 6, 80805 M\"unchen, Germany}

\maketitle 

\def\beq#1\eeq{\begin{equation}#1\end{equation}}
\def\beql#1#2\eeql{\begin{equation}\label{#1}#2\end{equation}}

\def\bea#1\eea{\begin{eqnarray}#1\end{eqnarray}}
\def\beal#1#2\eeal{\begin{eqnarray}\label{#1}#2\end{eqnarray}}

%\newcommand{\beq}{\begin{equation}}
%to include label
%\newcommand{\beql}[1]{\begin{equation}\label{#1}}
%\newcommand{\eeq}{\end{equation}}

%\newcommand{\beq}{\begin{equation}}
%\newcommand{\eeq}{\end{equation}}
%\newcommand{\bea}{\begin{eqnarray}}
%\newcommand{\eea}{\end{eqnarray}}
\newcommand{\Z}{{\mathbb Z}}
\newcommand{\N}{{\mathbb N}}
\newcommand{\C}{{\mathbb C}}
\newcommand{\Cs}{{\mathbb C}^{*}}
\newcommand{\R}{{\mathbb R}}
\newcommand{\intT}{\int_{[-\pi,\pi]^2}dt_1dt_2}
\newcommand{\cC}{{\mathcal C}}
\newcommand{\cI}{{\mathcal I}}
\newcommand{\cN}{{\mathcal N}}
\newcommand{\cE}{{\mathcal E}}
\newcommand{\cA}{{\mathcal A}}
\newcommand{\xdT}{\dot{{\bf x}}^T}
\newcommand{\bDe}{{\bf \Delta}}

\newcommand{\Eq}[1]{Eq\,\ref{#1}}
\newcommand{\tfrac}[2]{{\textstyle\frac{#1}{#2}}}
\newcommand{\ket}[1]{| #1 >}
\newcommand{\bra}[1]{< #1 |}
\newcommand{\ome}[2]{<#1|{\cal O}|#2>} 
\newcommand{\gme}[3]{<#1|#3|#2>}
\newcommand{\spr}[2]{<#1|#2>}
\newcommand{\eq}[1]{Eq\,\ref{#1}}

\def\dr{detector }
\def\drn{detector}
\def\dtn{detection }
\def\dtnn{detection}

\def\pho{photon }
\def\phon{photons}
\def\phos{photons }
\def\phosn{photons}
\def\mmt{measurement }
\def\ple{particle }
\def\plen{particle}
\def\an{amplitude}
\def\a{amplitude }
\def\co{coherence }
\def\con{coherence}

\def\st{state }
\def\stn{state}
\def\sts{states }
\def\stsn{states}

\def\cow{``Collapse of the Wavefunction"}
\def\de{decoherence }
\def\den{decoherence}
\def\dm{density matrix }
\def\dmn{density matrix}

\newcommand{\mop}{\cal O }
\newcommand{\dt}{{d\over dt}}
\def\qm{quantum mechanics }
\def\qms{quantum mechanics }
\def\qml{quantum mechanical }

\def\qmn{quantum mechanics}
\def\mmtn{measurement}
\def\pow{preparation of the wavefunction }

\def\me{ L.Stodolsky }
\def\T{temperature }
\def\Tn{temperature}
\def\t{time }
\def\tn{time}
\def\wfs{wavefunctions }
\def\wf{wavefunction }
\def\wfn{wavefunction} 
\def\wfsn{wavefunctions}
\def\wvp{wavepacket }
\def\pa{probability amplitude } 
\def\sy{system } 
\def\sys{systems }
\def\syn{system} 
\def\sysn{systems} 
\def\env{environment }
\def\envn{environment}
\def\ha{hamiltonian }
\def\han{hamiltonian}
\def\rh{$\rho$ }
\def\rhn{$\rho$}
\def\rhss{$\rho_{ss}$ }
\def\rhssn{$\rho_{ss}$}

\def\op{$\cal O$ }
\def\opn{$\cal O$}
\def\yy{energy }
\def\yyn{energy}
\def\pz{$\bf P$ }
\def\pzn{$\bf P$}

\def\plz{polarization  }
\def\plzs{polarizations }
\def\plzn{polarization}
\def\plzsn{polarizations}

\def\sctg{scattering }
\def\sctgn{scattering}
\def\env{environment }
\def\envn{environment}
\def\rn{random noise }
\def\rnn{random noise}
\def\ss{small system }
\def\ssn{small system}
\def\sgc{{\bf\sigma C}}
\def\sgb{{\bf\sigma B}}
\def\hf{\tfrac{1}{2}}
\def\om{\omega}
\def\bp{$\bf P\; $}
\def\bv{$\bf V\; $}
\def\phi{x}

\begin{abstract}
 We study the  effects of noise and decoherence for a
double-potential
well  system,
suitable for the  fabrication of qubits and quantum logic elements.
 A random noise
term is added to the  hamiltonian, the resulting \wf found
numerically and the  \dm  obtained by averaging over noise signals.
 Analytic solutions using the two-state model are obtained and 
found to be  generally in agreement 
with the numerical calculations.

   In particular,
 a simple formula for the \de rate in terms of the noise
parameters in the two-state model
 is reviewed  and  verified for the full
simulation with the multi-level \syn. The
formalism is
extended to describe multiple  sources of noise or different
``dephasing''
axes  at the same time. Furthermore, the old formula  for the
``Turing-Watched Pot'' effect is generalized to  the case where the
environmental
interactions do not conserve the ``quality'' in question.

 Various forms for the noise signal are investigated.
 An interesting result  
is the importance of the  noise power at
low frequency. If it vanishes there is, in
leading order, no \den. This is verified 
in a  numerical simulation where  two apparently similar noise
signals, but differing in the power 
at zero frequency, give  strikingly different \de effects. A short
discussion  of situations dominated by low frequency noise is
given.

\end{abstract}

\section{Introduction}
In the study of macroscopic \qm and quantum logic devices in
particular, the question of \de and its effects remains an
important, if not the most important, issue. 
Inevitable disturbances from the environment of the quantum device,
and correlations established with variables of the environment,
will limit the length of time for which the \sy under study acts as
an isolated quantum system.

 In previous work we studied, by means of numerical simulation,  
how  \de affects and interacts with 
the behavior of some quantum logic devices \cite{one}.
The \de was modeled by a simple random noise signal  presumed to
act on a certain component of the \sy--  
in the SQUID corresponding to flux noise.
 In the present work we wish to consider more generally the
connection between 
\den, the type of  random noise signal,  and the parts of the 
\sy upon which it acts. We reach a number of interesting
conclusions on the effects of these factors.

 A further interest of such studies is its connection with the 
``quantum measurement problem",  and we are able to illustrate
quantitatively phenomena like the \cow \, and the ``Watched Pot
Effect"
through the simulations.

\section{Hamiltonian}

 We shall examine the \ha
\begin{equation}\label{hama3}
H={-1\over 2\mu }{\partial ^2\over\partial \phi^2}
 +V_0\{\tfrac{1}{2}[(\phi-\phi^{ext})^2 ]+\beta\, cos\phi\}+H_{\cal
N}(t)\; ,
\end{equation}
which is suitable for representing a "qubit" and through
manipulations on $\phi^{ext}$, the quantum logic operations NOT and
CNOT \cite{one}.

\begin{figure}[htbp]
\begin{minipage}{0.49\textwidth}
  \includegraphics[width=1.0\textwidth]{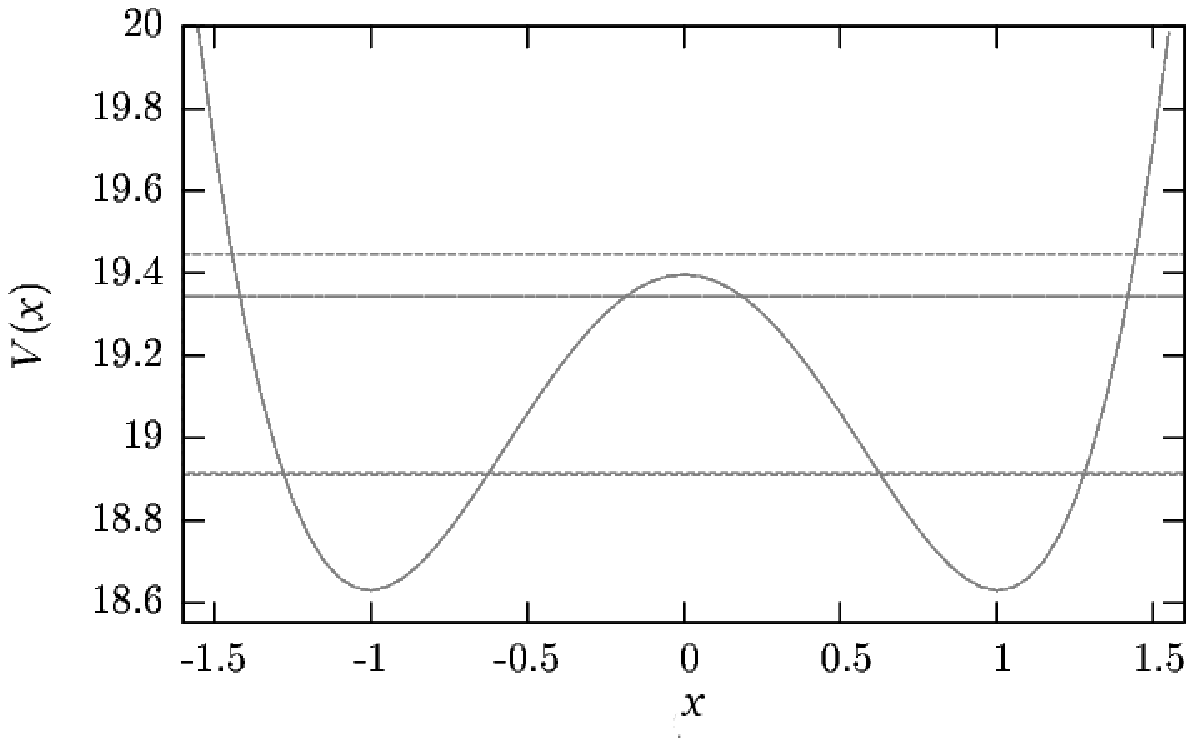}
\end{minipage}
\hfill
\begin{minipage}{0.49\textwidth}
  \includegraphics[width=1.0\textwidth]{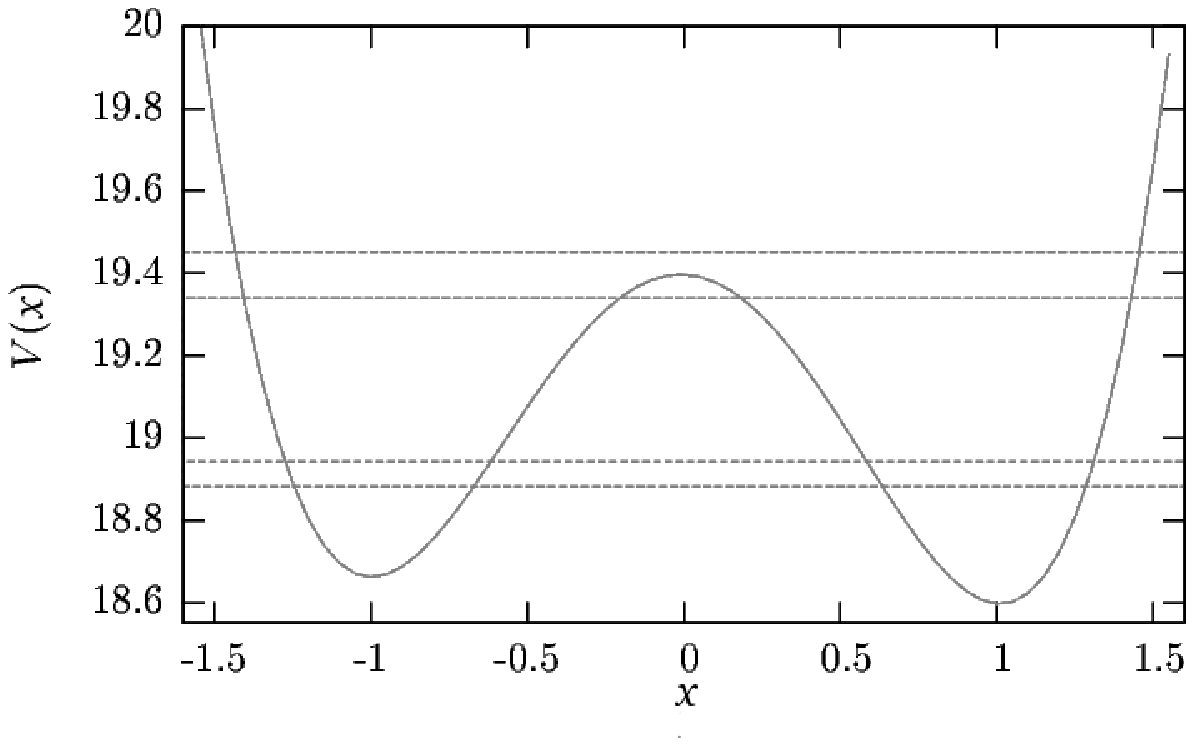}
\end{minipage}
\caption{Double well potentials ($V_o$  term) of the \ha
\eq{hama3}, with the
first four
energy levels indicated. 
  The left panel shows the symmetric   
configuration with $\phi^{ext}=0$ while the right one shows an
asymmetric configuration with $\phi^{ext}=0.0020$. The behavior of
the two lowest
levels is well represented by an effective \ha $H=\hf\bf \sigma
V$, where $V_x$ is given by the tunneling \yy and $V_z$ by the
asymmetry of the wells. }\label{wells}
\end{figure}

The first part of the expression
represents a double potential well problem illustrated by Fig\,
\ref{wells}. The $H_{\cal N}(t)$ represents a small, time
dependent, random noise term. This will be used to simulate the 
\de effects.  These could be due to either actual noise sources in
the laboratory like stray fields, or true \qml \de due to
unavoidable
interactions of the simple idealized \sy with 
other variables.

  The  parameter $\phi^{ext}$ controls the
asymmetry of the two potential wells, $\phi^{ext}=0$ being the
completely symmetric situation where the  \yy splitting between
the two lowest levels $\omega_{tunnel}$ is only due to the
tunneling
through the barrier.
The height and width of the barrier are controlled by $\beta$.
 In  the previous work it was found
that the values  $\mu=V_o=16.3;\ \,\beta=1.19$  give
 behavior suitable for quantum logic elements and we shall use
these values in our simulations here. As seen on the figure, with
these values there is a well defined pair of levels below the
barrier. At the same time the barrier is  low, giving a  relatively
large tunnel splitting. This permits the adiabatic sweeps, on which
the gate operations are based, to be fast.
For these values one has
 $\omega_{tunnel}=0.0044$.

For the SQUID, these parameters are related to the inductance L and
capacitance C by  $\mu=V_o\approx 1030 \sqrt{\frac{C/pF}{L/pH}}$.
The parameter $\beta$ characterizes the critical current $I_c$ and
is given by $\beta=\frac{2\pi L I_c}{\Phi_o}$.  (We use $\hbar,c=1$
units.) The overall energy scale is given by $E_o=1/\sqrt{LC}$ and
the time unit by $1/E_o=\sqrt{LC}$. Thus the time unit is 
$1.0\times 10^{-12} seconds\times (L/pH
~C/pF)^{1/2}$. With  $L=400 pH$ and $C=0.1pF$ for example,  
$\omega_{tunnel}=0.0044$
corresponds to  $\sim 4\times 10^{-7}eV$. The time unit is then
$6.3\times 10^{-12}~sec$.
\cite{units}.

 The numerical methods are as outlined in \cite{one}, as introduced
in \cite{jac}. A large basis of harmonic oscillator states is
employed, and the resulting sparse \ha matrix is inverted to find
the \yy eigenstates. The time evolution is found by repeated
application of ${(1+\hf i H\Delta t)}/{(1-\hf i H\Delta t)}$.  

\section{Two-State System}\label{sec3}

 The previous studies have shown that there is a range of parameter
values where the two
lowest levels for \eq{hama3}
can be adequately represented as an effective two level \sy
with \ha $H=\frac{1}{2}{\bf \sigma V}$.

 We  use an (x,y,z) coordinate system for the
various vectors in the following way. $\bf V$ is approximately in
the `z' or `up'
direction when $\phi^{ext}$ is relatively large so that the
potential is asymmetric and the \yy level splitting is essentially
determined by the distance between the bottoms of the potential
wells. $\bf V$ is in the `x' direction or `horizontal' when 
$\phi^{ext}=0$ and the \yy splitting is determined only by the
tunneling
through the barrier. Specifically, one determines \cite{one}  the
components of \bv  to be $V_z \approx 2 V_o
\phi^{ext}
\phi_c$ and $V_x=\omega_{tunnel}$, the tunneling \yyn. The
quantity $\phi_c$ is the value of the x-coordinate where 
  the \wf is centered when the \wf is localized in just one of
the potential wells. For the parameter values used here this is
always close to
1, about 0.9. The value of 
  $\omega_{tunnel}$, namely =0.0044,
is  obtained from the
numerical solutions  as the level splitting for $\phi^{ext}=0$,
with the chosen parameters $\mu=V_o=16.3;\ \,\beta=1.19$.

According to \cite{one}, deviations from the two-state model begin
to appear when the level
splitting for the lowest pair becomes comparable to the distance to
the
next set of levels. Various tests  showed that the
effective two-state description was good up to about 
$\phi^{ext}\approx 0.01$, where the level splitting was $ 0.30$.

To represent the noise 
 in this approximate two-state picture, we assume an  additional
term  $\hf{ \sgb} $ in  the two-state \ha
with a  time
dependent random field $\bf B$.  The effective total
 two-state \ha is then
\beql{tst}
H=\tfrac{1}{2}{\bf \sigma V}+\hf{ \sgb}\,.
\eeql
 $\bf B$  thus represents a rapid
small stochastic addition \cite{bnorm} to the $\bf V$. Just as for
the components of $\bf V$, it is possible to establish  relations
between the $\bf B$ and the parameters of $H_{\cal N}$. These will
be given in section \ref{rltn}.

\section{Density Matrix}
 Our object of study is the \dm  of the  \syn.
 In our random noise model we  generate it by inserting a
particular realization 
${\cal N}^a(t)$ of a noise signal in the \ha \eq{hama3} and
then evolving  an
initial \wf to a final time with this \ha $H^a$. Repeating the
process,  and  averaging 
over many different noise realizations, N, gives a final \dmn :
\begin{equation} \label{av}
\rho(t)=\overline{\psi \psi^{\dagger}}={1\over N}\sum_{a=1}^N
{\psi^a(t)
\psi^{a\dagger}(t)}\; .
\end{equation}
Although
the
evolution is  unitary for a given ${\cal N}^a(t)$, the averaging
process is not and the procedure evidently violates  unitarity. The
resulting \dm
will thus
exhibit a time-dependent \den, which we  wish to examine.

In particular  we  would like to study the \de for the effective
two-level \sy comprised of the lowest states of the \han, with its
$2\times 2$ \dmn. However,
for the calculations with the full \eq{hama3}, it must be noted
that the resulting numerical \dm \eq{av}   is a function of two
continuous variables in $\rho(\phi',\phi)$. Thus in principle
\eq{av} refers not
only to the two lowest states of the \han, but to the entire
Hilbert space of the continuous $\phi$ variable.
 
Nevertheless, in the previous studies  
it was found that in the region of parameter space where the
two-state description is valid,  the Hilbert space of lowest states
is accurately spanned by one and the same set of \wfs  even as the
parameters of the \ha are varied and the \yy eigenstates change
\cite{cmplt}.  Therefore \wfs beginning in the two-state space 
remain there, and  the results of evolving  with \eq{hama3}  can 
be expressed in the two-state space by projecting them onto the
corresponding pair of \wfsn. This is the procedure we will follow
in the numerical evaluation of the $2\times 2$ \dmn.

On the other hand, for the simplified  
 two-state or spin 1/2  \sy represented  by \eq{tst}, the
 \dm can be predicted
analytically, as will be explained in section \ref{anly}.  This
analytical result may be compared with that for the numerical
procedure just described,  where  we insert the noise signal in 
\eq{hama3}.   To carry
out this comparison one needs a translation between
the noise signal inserted  in $H_{\cal N}$ and the equivalent $\bf
B$ in \eq{tst}. This translation will be given in section
\ref{rltn}.

The two-state \dm  can be parameterized in terms of Pauli matrices
as
\begin{equation}\label{rhomata}
\rho= \tfrac{1}{2} (1+{\bf P}\cdot {\bf \sigma}) \; .
\end{equation}
With  our coordinate conventions $\bf P$ is in the ``up" or
``down"
or $\pm z$ direction when the wavefunction is localized in one of
the
potential wells of Fig\,\ref{wells}; and it is in the
``horizontal''
or $\pm$ x direction when  the \wf is the symmetric or
antisymmetric combination  of states in both potential wells.
$|{\bf P}|=P=1$
corresponds to pure state and $P<1$ a mixed state.

In the two-state model the ``polarization'' vector $\bf P$
evolves according to \cite{us}
 
\begin{equation}\label{pdota}
\dot {\bf P}={\bf P}\times {\bf V}- D\,{\bf P}_{T}\; ,
\end{equation}
where the $-D\,{\bf P}_{T}$ term represents the quantum damping or
loss
of coherence. $D$ is the \de rate and $D^{-1}$ is the \de time.
This term reduces the components $\bf P$ 
transverse to the direction chosen by the noise or \de
interaction.

Although
in the present paper we concentrate on situations with constant
parameters in the \ha (aside from the noise term),  in   previous
work time dependent \bv 's   
 were used. These may be employed to realize adiabatic  quantum
logic gates
\cite{one}, and via the adiabatic method to give a  direct
measurement of $D$ by a turning on-and-off of
 the classical-quantum transition \cite {oof}.

\section{D in the two-state model}\label{anly}

We now consider how the \de or
damping and the associated  parameter $D$ are generated by the
random noise
in the two-state picture of \eq{tst}. 
 The $\bf V$
 are assumed 
constant or slowly varying over the time period necessary to find
the decoherence
effects induced by the random fields.
 The method is essentially the
same as in Ref\,\cite{one} but we would like
to   present it in a somewhat more general form.
      This is nicely done by
following Ref\,\cite{kh}. One considers  the \dm
$\rho=\tfrac{1}{2}(1+{\bf P\cdot\sigma})$ in the Heisenberg
representation. The $\bf P$ are also assumed to be slowly varying
so that at a given time
one may assume a certain initial \dm which we call $\rho(0)$.

In \eq{av} one may denote each of the terms on the right as $\rho^a
$, so that $\rho={\overline \rho{^a}}$.
  We
then examine  the \dm
equation $\dot \rho=-i[H(t),\rho]$ term by term
\beql{hl}
\dot \rho^a=-i[H^a(t),\rho^a] \;,
\eeql
iteratively: 
 $\rho(t)\approx \rho(0)+\rho_1+\rho_2$. The different orders refer
to how many factors of the small $\bf B$ occur   and one compares
equations with the same number of $B$'s \cite{bfact}.
Since we will assume the average
value of the random fields $\overline{\bf B}$ to be  set to zero
(or incorporated in the main \han), the first order contribution to
the average $\dot \rho$
is zero: ${\dot\rho_1}=-i[\overline{\sgb},\rho(0)]=0 $. We
thus
consider  the second  order
for $\dot \rho$

\beql{scn}
\dot\rho^a_2(t)=-i\hf[{\bf \sigma B}^a,
\rho^a_1]=-\tfrac{1}{4}[{\bf \sigma B}^a,[{\bf \sigma
C}^a,\rho(0)]\,]
\eeql
 where we introduce the shorthand  ${\bf C}(t)=\int_0^t dt'{\bf
B}(t')$ and use $\rho_1=-i\hf [\sgc, \rho(0)]$. 

Now in performing  the average over the random fields in \eq{scn},
 we will have to do with bilinear correlators between the vector
components 
 of $B,C$. These
 can be
thought of as
tensors of the type $\overline {X_iY_j}$, where the average
is over different realizations of the random fields. With just one
random field in a fixed direction $\bf e$, one  has
\begin{equation}\label{rnd1}
{\bf \overline {X\otimes Y}} = {\cal
C}\,{\bf
e}\otimes {\bf e}^{\dagger}
\end{equation}
for the tensor, where ${\cal C}$
represents some
quadratic correlator.

 By  making repeated use
 \footnote{This is the first point where
one uses the properties of the $\sigma$ as Pauli matrices.
Otherwise they could have been any  group generators, in which case
the cross product formulas are replaced by bilinear products of
structure constants \cite{kh}}  of
$[{\bf\sigma X},{\bf\sigma Y}]=i2{\bf \sigma
X\times Y}$  and comparing the resulting coefficients of $\sigma$,
 \eq{scn} implies 
\beql{dpt}
{\bf \dot P }=\overline{BC} ({\bf
e\times (e \times P)})\;.
\eeql
To evaluate the correlator $\overline{BC}$ we use
the basic result from the theory of stationary random noise that
the quantity
$\overline{(\int_0^tB(t')dt')^2}= 2t {\cal A}$, where ${\cal A}$ is
 the time integral of the autocorrelation
function
\beql{defa}    
{\cal A}=\int_0^{\infty}\overline{B(0)B(t)}dt\;.
\eeql
Then
\beql{cor}
\overline{BC}=\overline {B(t)\int_0^tB(t')dt'}=\tfrac{1}{2}
\frac{d}{dt}\overline{\biggl(\int_0^tB(t')dt'\biggr)^2}={\cal A}\;.
\eeql 

As is evident when taking the scalar product with $\bf e$, the
cross product expression
represents the component of $\bf P$ perpendicular to the respective
$\bf e$,
i.e. with the component along $\bf e$ removed. So we may call 
$( {\bf (eP)e- P})$= $\bf- P_T$ with ``T'' for transverse.

We therefore obtain that the D in \eq{pdota} is given by
\beql{d1}
D={\cal A}
\eeql
\eq{d1} has the typical form of a dissipative parameter related to
the integral over the autocorrelation function for a fluctuating
quantity\cite{chand}.

One can also handle more than one noise source by this method. 
With two independent random fields,  along
the fixed
directions $\bf e_1$ and $\bf e_2$, we will have a tensor of the
form
\beql{two}
{\bf \overline {X\otimes Y}}= {\cal C}_1{\bf e_1}\otimes {\bf
e_1}^{\dagger}+{\cal C}_2{\bf
e_2}\otimes
{\bf e_2}^{\dagger}\; .
\eeql
That is,   different noise effects can be added up
separately and do not interfere with each other. This is due to the
assumed
independence of the two noise signals. Note the
$\bf e_1$ and $\bf e_2$
need not be orthogonal.

Thus  according to \eq{two}, for  the case of two independent noise
signals
 \eq{dpt} generalizes to
\beql{dptb}
{\bf \dot P }= \bigl[{\cal A}_1
( {\bf (e_1P)e_1- P})+
 {\cal A}_2( {\bf (e_2P)e_2- P}) )\bigr]\;.
\eeql

Calling ${\bf (e_1P)e_1- P  =- P_{T1}}$,and ${\bf (e_2P)e_2- P  =-
P_{T2}}$
 we thus find that the old \cite{us} rotation- \de equation
\eq{pdota}
now becomes
\begin{equation}\label{pdotb}
\dot {\bf P}={\bf P}\times {\bf V}- D_1\,{\bf P}_{T1}- D_2\,{\bf
P}_{T2}\; ,
\end{equation}
with 
\beql{d}
D_1=  {\cal A}_1~~~~~~~~~~~~~~~~~~~~~~~~~~D_2={\cal A}_2
\eeql
Concerning dimensions,  $\bf B$ is an \yy in view of its role in
the \han, and so ${\cal A}$
via \eq{defa} is also an \yyn, which matches the units of $D$.    

The conclusion \eq{pdotb} could also have been arrived by an
intuitive
argument where one imagines turning off one noise signal 
for a short 
time and then repeating the process with the  other noise signal
turned
off. One would then obtain first the results of \eq{pdota} with
respect to
one  direction and then with respect to the other direction. The
net result would correspond to \eq{pdotb}. 

As said, the  $\bf e$'s can be in any direction, not necessarily
orthogonal; and   since the independent addition of signals in
\eq{two} holds for any
number
of different signals, evidently \eq{pdotb}  can be extended to more
than two noise signals
  by simply adding a $-D\,{\bf P}_{T}$ term for each. It is of
course
essential that the different noise signals be uncorrelated.

The length of $\bf P$ is always decreasing with an equation of the
type \eq{pdotb}  and these
considerations can  be given an interpretation in terms of the
increase of entropy \cite{ent}.
 In particular we  note a formula for the  
decrease of the length of $\bf P$ (squared):
\begin{equation}\label{pdotc}
\tfrac{1}{2}\frac{d}{dt}{\bf P}^2={\bf P}\cdot \dot{\bf P}=-
{\bf P}\cdot( D_1 {\bf P_{T1}}+D_2 {\bf P_{T2}})=-D_1 {\bf
P_{T1}}^2-D_2 {\bf P_{T2}}^2\; ,
\end{equation}
 for the case of two independent
\de or noise signals as in \eq{pdotb}.

Concerning the necessity of considerating more than one noise
source, it is
 possible or indeed likely that there are noise and \de 
effects on more than one aspect or component of a  \syn. For the
SQUID, for example, there might be
external flux noise  giving fluctuations in $\phi^{ext}$ and at the
same time
effects in the Josephson circuit amounting to fluctuations in
$\beta$. Such a  situation would be described by \eq{pdotb}.

With only one noise ``axis'' the \de problem may be viewed as one
of random phases or rotations around that axis, as was done in
\cite{one} for flux noise in the SQUID. This is sometimes 
called ``dephasing'', and $D$ is the associated diffusion
parameter. Thus it may be said that  \eq{pdotb}  gives the formula
for the evolution when there is more than one 
``dephasing''axis.

\section{Noise Models} \label{nm}

We shall  consider different types of noise signals $\cal N$(t).
Noise signals
are characterized by their power spectrum ${\cal P}(\om)$, which is
the fourier transform of the autocorrelation function
$\overline{{\cal
N}(0){\cal N}(t)}$, namely ${\cal P}(\om)=\int_0^{\infty} dt\,
cos\om t\,\overline{{\cal
N}(0){\cal N}(t)}$. One such power spectrum is  ``white noise''
with a cutoff  $\om_c$:
\beql{n1}
{\cal P}(\om)={\om_c^2\over \om^2+ \om_c^2}\;.
\eeql 
This is  a constant or white noise spectrum up to 
frequency $\om_c$, and  then falls off as $1/\om^2$ at high
frequency.
This was used in a simplified way in \cite{one}.
 In  section\,\ref{vln}  below we will discuss some examples using
this
kind of noise.

In addition,  we 
also use
planckian spectra  corresponding to 1, 2,  or 3 spatial dimensions.
With these we can to try to represent a thermal background and as
will be explained below, allows us to conveniently compare cases
with and without noise power at zero frequency. They also allow a
simple variation of the frequency content by changing $T$
which is a ``\T''
parameter characterizing the frequency spectrum, 
with an exponential cutoff above $\om \sim T$. 
For n=1 we  take the planckian form
\beql{np}
{\cal P}(\om)= {(\om/T)\over e^{\om/T}-1}
~~~~~~~~~~~~~~~~~n=1\;.
\eeql 
 
We have fixed the normalizations in \eq{np} and \eq{n1}  
such that  ${\cal
P}(0)=1$.  
  This condition corresponds to  
 setting the time integral of the autocorrelation function  
equal to one:
\beql{nn}
{\cal P}(0)=
\int_o^{\infty} dt \,\overline{{\cal
N}(0){\cal N}(t)}= {1\over N}\Sigma_a \int_o^{\infty}dt\,{\cal
N}^a(0){\cal N}^a(t)=1\;,
\eeql
where ${\cal N}^a(t)$ is a particular realization ``a'' of the
signal, and we have generated $N$ such signals.
This convention is particularly convenient since according to
\eq{d1} the \de
parameter $D$ is simply proportional to just this time integral of
the
autocorrelation  function.
 With this
standardized normalization of the noise signal we will 
regulate  its overall strength in different applications
 by an adjustable small coupling factor.

The planckian forms corresponding to spatial dimensions n=2 and n=3
cannot be normalized in this way since they contain extra
powers of $\om$, giving  vanishing noise power at $\om =0$. We
therefore choose to normalize them such that the variance of the
signals,  which is a measure of their magnitude,
 is equal to that for n=1 at the same T. 

This leads to the following definitions
\bea \label{twth}
{\cal P}(\om)&= &{\pi^2\over 12 \,\zeta(3)}\;{(\om/T)^2\over
e^{\om/T}-1}
~~~~~~~~~~~~~~~~~n=2 \cr
&= &{5\over 2\pi^2}\;{(\om/T)^3\over e^{\om/T}-1}
~~~~~~~~~~~~~~~~~~~~~n=3
\;.
\eea

The variance of the signal corresponds to the integral over the
power spectrum, so with these normalizations the integral
$\int_0^{\infty}d\om\cal{P}(\om)$ is the same for all three signals
 in \eq{np} and \eq{twth}.

Specifically, one finds

\beq \label{tref} 
  \overline{{\cal N}^2}=
\frac{2}{\pi}\int_0^{\infty}d\om\,{\cal{P}}(\om)
= T\frac{\pi}{3}
\eeq 

For T=0.025, which  we shall use in many of the simulations, this
gives  a typical signal strength of
$\surd \overline{{\cal N}^2}=0.16$. 
As we shall see below in section \ref{pw},
the case of  noise signals with the property
$\int_o^{\infty}dt\overline{{\cal
N}(0){\cal N}(t)}=0$, as in \eq{twth}, are of special interest.

To numerically  generate a particular signal ${\cal N}^a(t)$ as a
realization of
one of
the above models, we work with a discretized  fourier space where 
${\cal N}(t) =\Sigma_{\om} {\cal N}_{\om}e^{i\om t}$. A given
random set $a$ of the ${\cal N}_{\om}$ then determines the signal.
We
have used two procedure to generate the ${\cal N}_{\om}$.

In the random phase procedure, one takes $\surd {\cal P}(\om)$
times
a random phase $e^{i\theta}$ with $\theta$ chosen randomly and
uniformly in the interval  
$0\geq \theta\geq 2\pi$.

 In the gaussian modulus method one chooses
a real and imaginary part of ${\cal N}_{\om}$ randomly from a
gaussian distribution with zero mean and variance  $ {\cal
P}(\om)$.
 Both methods lead to an ensemble with the power spectrum  $\sim
{\cal P}(\om)$. 
We find that both procedures lead to very similar results in our
simulations.

\begin{figure}[h]
\includegraphics[angle=-90,width=\hsize]{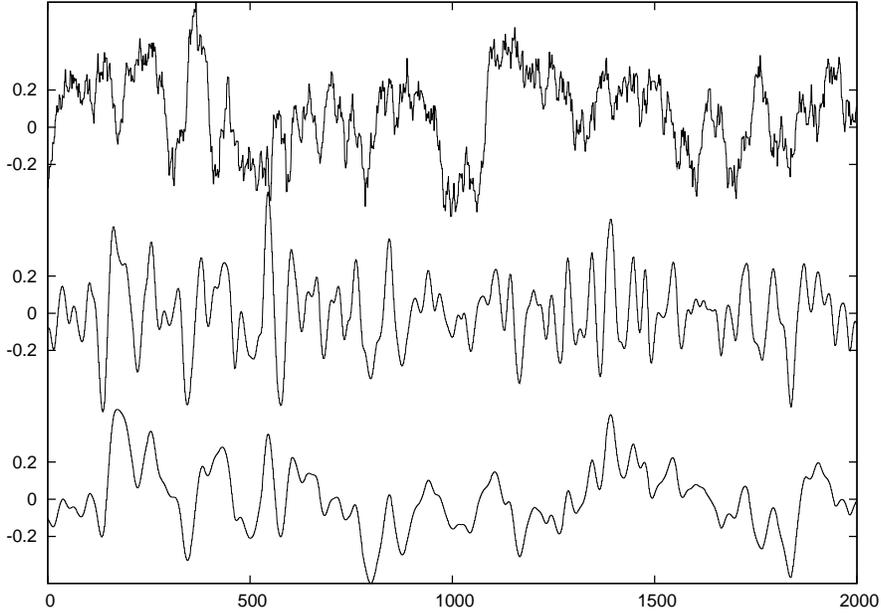}
\caption{Samples of  noise signals ${\cal N}(t)$. Upper curve:
Cut-off white noise from
\eq{n1} with
cutoff $\om_c=0.025$. Middle curve: planckian noise for n=3 as in
\eq{twth} with T=0.025.  Lower curve:  planckian noise for n=1 as
in \eq{np} with T=0.025. 
 The signals are normalized to have the same
variance or integrated noise power but differ in having
zero (middle) or non-zero (upper and lower) noise power at zero
frequency. The high frequency tail present in the power spectrum 
of the
upper, non-planckian, curve is evident. With  $L=400 pH$ and
$C=0.1pF$ for the SQUID, the time unit in  this and all further
plots would be $6.3\times 10^{-12}~sec$.}
\label{samp}
\end{figure}

Figure \ref{samp} shows some samples of the different noise
signals. The upper curve is for \eq{n1}. One sees the presence of
the high frequency tail in the power spectrum. The middle curve
is for the planckian n=3, \eq{twth}, and the lower curve
 for the planckian  n=1, \eq{np}. These latter two differ in that
for n=3 there is no power at zero frequency, while for n=1  it is
present. The temperature T=0.025 or $\omega_c$ is the same for all
three. It will be seen that the amplitude of the signals is roughly
in accord with \eq{tref}, with $\surd \overline{{\cal N}^2}=0.16$.

\section{Relation Between $H_{\cal N}$ and  $\bf B$} \label{rltn}

In this section we determine the relations between the random $\bf
B$ of the two-state description \eq{tst} and the noise term in the
full \ha
\eq{hama3}. Given these relations, we will be able to compare the
results of a full numerical simulation with \eq{hama3} and the
analytical predictions from \eq{d1}.

As explained below,  noise in
the `z' direction arises through  fluctuations in $\phi^{ext}$, 
and in the `x' direction, fluctuations in $\beta$. In the case of
the SQUID these correspond to noise/\de in the external flux  and 
 in the junction circuit,   respectively. It is quite plausible
that  they should  correspond to independent noise sources.

\subsection{ Noise in the ``z-direction''} \label{sz} 
We thus consider adding a noise term representing a fluctuation of
$\phi^{ext}$ in \eq{hama3}: 
\beql{znoi}
H_{\cal N}=  \,\delta \phi^{ext} V_o \, \phi= \,\delta \eta \,
{\cal N}(t) V_o  \, \phi \; .
\eeql
 We call the fluctuation $\delta \phi^{ext}$ and for a particular
noise signal "a" it is represented by $\delta \eta \,
{\cal N}^a(t)$ with
$\delta \eta$  an overall coupling strength.
By using  the identification $\phi\approx \phi_c \sigma_z $
(see section \ref{sec3}) connecting the operator $x$ of the full
\ha with the $\sigma$ of
the spin 1/2 picture, on sees   that 
 this $H_{\cal N}$ is equivalent \cite{bnorm} to
the B field of the two-state model as :
\beql{bext}
B_z=2 V_0 \phi_c \,\delta \eta {\cal N}\; ,
\eeql
with other B components zero.
We therefore predict, according to \eq{d1}, that when using a noise
signal  normalized with ${\cal P}(0)=1$ as explained in section
\ref{nm},   a \de
constant from the full simulation with \eq{hama3}
\beql{znoise}
 D=\int_0^{\infty}\overline{B_z(0)B_z(t)}dt=4 (V_0 \phi_c \delta
\eta)^2=0.89\times 10^3\; (\delta \eta)^2\;,
\eeql
where in the last step  we  insert 
$V_0 =16.3, \phi_c=0.9$.

\subsection{Noise in the ``x-direction"} \label{subx}

Switching to the ``x-direction",
 we consider the effect of  small variations of $\beta$ in
\eq{hama3}, amounting to fluctuations in the barrier potential.
Since  $\beta$ regulates the  tunneling,  a change
in  $\beta$ induces a change in the tunneling \yyn.
Thus in the two-state \ha \eq{tst} a variation in $\beta$ induces 
(apart from an unimportant constant shift) a change in the
$\sigma_x$ term, a $B_x$. Other components of B are zero.

To obtain the magnitude of  $B_x$  \cite{note},
one  notes that the coefficient of $\sigma_x$ is the tunnel
splitting;
 $V_x+B_x=\omega_{tunnel}+\delta\omega_{tunnel}$. Thus   
$B_x=\delta\omega_{tunnel}=\delta
\beta\,\frac{d\omega_{tunnel}}{d\beta }$.

\begin{figure}[h]
\includegraphics[width=0.8\hsize]{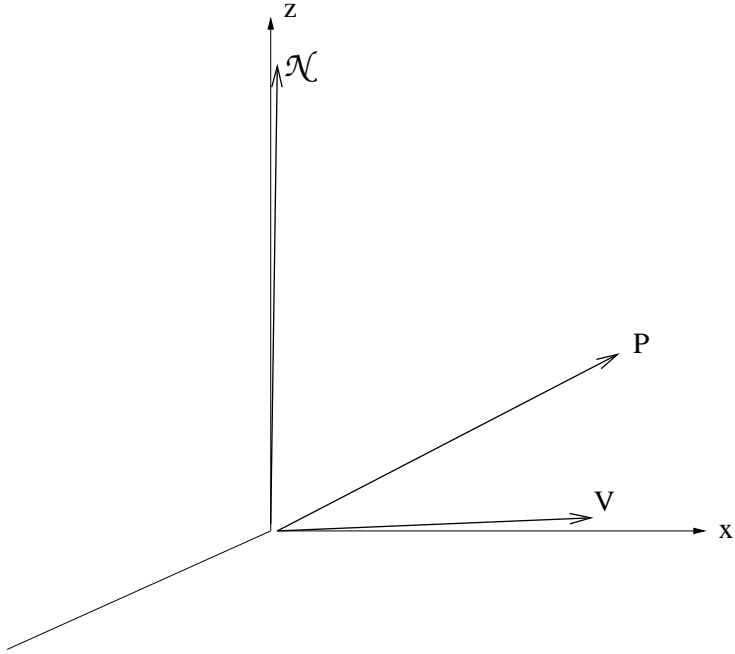}
\caption{Typical configuration of  vectors used in the
simulations. With no damping the polarization vector 
 \bp rotates around \bv, so with \bv in the x-direction there are
oscillating components $P_z, P_y$.
 For D not too large these are damped oscillations
 as the component of \bp   perpendicular to the
noise direction $\cal N$ or $\bf B$ is reduced. For very strong
$D$, \bp is ``pinned'' along the noise vector, corresponding to the
\cow.
The initial \bp is of unit length and along the z-axis
for Figs\,\ref{znoilr} and \ref{xrlx}, while for Fig\,\ref{wp}  it
lies at 45 degrees
in the x-z plane.}
\label{vpn}
\end{figure}

Although there are analytic methods of evaluating tunneling
energies,
 they involve  exponentially
sensitive  effects and it seems best to find
$\frac{d\omega_{tunnel}}{d\beta }$ from our previous 
numerical results. In the vicinity of $\beta=1.19$, (and with
$V_0=\mu=16.3$ where we 
carry out our simulations)  examination of Table II of \cite{one}
 yields $\frac{d\omega_{tunnel}}{d\beta }\approx -0.18$ near
$\beta=1.19$.
Thus, for fluctuations  around $\beta=1.19$ the $\bf  B$ field
associated
with a change in $\beta$  is
\beql{bx}
B_x= \delta \beta\,\frac{d\omega_{tunnel}}{d\beta }=(-0.18){\delta
\beta}=(-0.18)\delta \beta_o \,{\cal N}\;,
\eeql
 We write the fluctuations in $\beta$ as $\delta \beta=\delta
\beta_o \,{\cal N}$, with ${\cal N}$ one of our noise signals and 
$\delta \beta_0$ a strength parameter.
 According to \eq{d1}, when ${\cal N}$ has the integral of the
autocorrelation function normalized to one,  
\beql{flba}
D=0.032 \;\delta \beta_o^2\,.
\eeql

For the numerical calculations, we wish to insert the same $\delta
\beta$ fluctuations in \eq{hama3}.
Since $\phi^2/2$ is
the leading term in $cos\,\phi$ which contributes to the level
splitting, a small variation in  $\beta$  can
be adequately represented by adding a term
$\delta \beta\,V_o \phi^2/2$ in the full \han .    We thus should
set, in 
\eq{hama3} 
\beql{add}
H_{\cal N}=-{\delta \beta}\,V_o \; \tfrac{1}{2} \phi^2= -{\delta
\beta}_o\, {\cal
N}\,V_o \; \tfrac{1}{2} \phi^2\, .
\eeql

\subsection{Comparison of x and z-noise} \label{xzcomp}

Our two examples of `z-noise' and `x-noise' are instructive in that
 the effective B fields for
the two-state \sy \eq{bext} and \eq{bx} and the associated D's,
\eq{znoise} and \eq{flba} differ widely for similar
  noise signals. The 
 difference originates  in  the fact that the 
`z-noise' operator is linear $\sim x$, while the `x-noise' operator
is quadratic $\sim x^2$.  The two lowest states comprising our
two-state \sy  have, with $x^{ext}=0$, odd or even parity. Thus
they will be strongly connected by the parity odd $x$ operator. For
the 
 $ x^2$ operator on the other hand  the change in the \yy
splitting is only due to the different shifts of the two states
under $ x^2$ and so is much smaller. For  $x^{ext}\neq 0$
but small the situation will be qualitatively similar.

Thus the symmetry properties associated with the noise or \de 
can be important. With specific devices this may
suggest a choice of operating conditions where the effects can be
minimized.

\begin{figure}[h]
\includegraphics[angle=-90,width=\hsize]{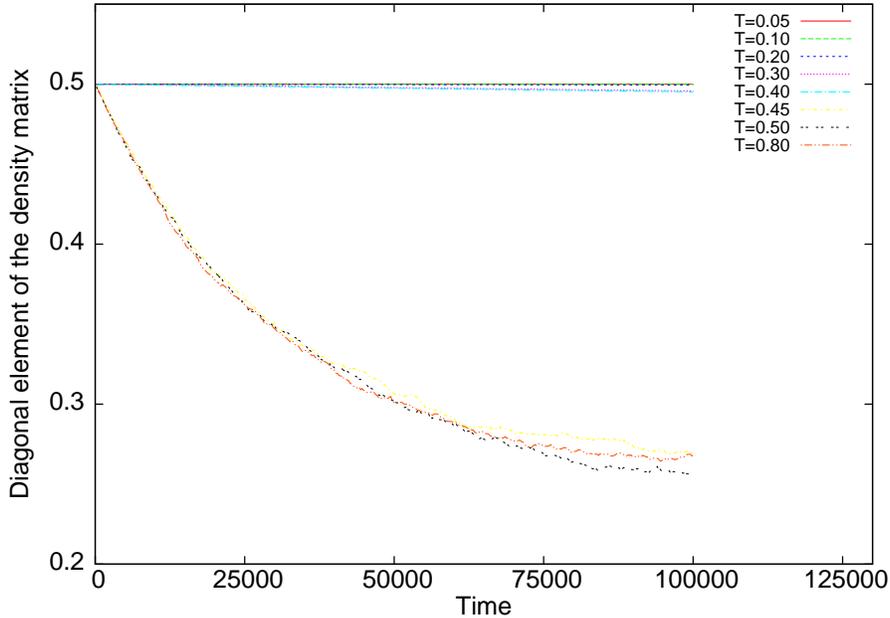}
\caption{Excitation to higher states.  When the noise power
spectrum
contains frequencies above that of the principal level spacing,
here
0.4, the diagonal elements of the 2x2 \dm fall below 1/2. When it
does not contain these frequencies the relaxation is to 1/2.}
\label{rlx}
\end{figure}

\section{Tests by Numerical Simulation}
We  turn to numerical tests of the formalism and the above results
\eq{znoise} and
\eq{flba}  of  the 
spin-$\hf$ picture. We check them against results from the full \ha
\eq{hama3} with the
corresponding $H_{\cal N}$, namely \eq{znoi} or \eq{add},  inserted
in the \han. We  find the \de parameter D from a fit to the damping
of $|\bf P|$ in the numerical calculations and
compare it with the analytic predictions.
For our various tests we chose a
coupling $\delta \eta $ such that the \de time 1/D is on the order
of some thousands of time units. For the typical SQUID parameters
of mentioned in the introduction this corresponds to a \de time in
the tens of nanoseconds.

\subsection{  Parameter Ranges }\label{hlf}
There will be certain conditions on the  range  of parameters where
the  analytical arguments  can be applied.

 One condition concerns large
perturbations, where the nature of the \sy is changed
substantially. These are evidently beyond the scope of the
theoretical method, and would
have to be dealt with simply by simulation alone. This is
particularly relevant in our example of $\beta$ fluctuations or `x-
noise' where the tunneling \yy is very small and $\beta$
fluctuations have a relatively large effect. This is discussed
in sect.\ref{x-n}.

A second condition concerns the frequency spectrum of the noise, as
characterized by the T or the $\om_c$ of the noise power spectra.
If the
dominant noise frequencies are too low, on the order of, or less
than the
important frequencies of the \syn,  this is in contradiction to our
assumption that the \sy can be taken as constant over many cycles
of the noise \cite{low}. In sect\,\ref{vln} we shall discuss some
features of low frequency noise.    On the
other hand, if the
frequencies in the noise are too high,  one risks exciting higher
states and so violating  our assumption of an effective two-state
\syn.
Since the important \sy frequency we   deal with is
$\om_{tunnel}= 0.0044$, and the
distance to the next set of levels is
about $0.4$, one  has the condition $0.0044 <<T,\om_c<<0.4$.
This criterion applies of course for our typical configurations and
 would have to be adjusted for others.

The excitation to higher states above that of our effective two-
level \sy
can be detected in that the diagonal elements of the 2x2 \dm will
relax not to 1/2, but to some smaller value, reflecting the missing
probability. We can perform an amusing numerical experiment on this
by using a `rectangular' noise power spectrum ${\cal P}(\om) $
which is exactly zero above some $\om=T$. The resulting behavior of
the \dm is shown in Fig \ref{rlx}
for various values of T.

\begin{table}
\begin{center}
\begin{tabular}{|l|l|l|} 
\hline
$\bf V$&$\bf P_{initial}$&$D$~~~\\
\hline
\hline
$(0.0044,0.0,0.0)$&(1,0,0)&0.00134\\
\hline
$(0.0044,0.0,0.0)$&(0,0,1)&0.00122\\
\hline
(0.0044,0.0,0.006)&$(1,0.0,-1){1\over\surd2}$&0.00118\\
\hline
(0.0044,0.0,0.02)&(1,0,0)&0.00141\\
\hline
\end{tabular}
\end{center}
\caption{Study of the constancy of D for various starting
conditions. A n=1 planckian noise in the `z-direction' with
$T=0.05$ is used. The coupling is $\delta \eta=0.00128$, for which
the prediction \eq{znoise} is $D=0.00146$. In the first two entries
\bv is along the x-axis while the starting \bp is first along the
x-axis and in the second line along the z-axis. In the third entry
\bv is close to 45
degrees with the z-axis and \bp starts orthogonal to it in  the
lower hemisphere. In the last and fourth entry \bv makes an angle
of only 12 degrees with the z-axis and \bp starts
along the x-axis.   One observes that  at about the 20\% level  the
value of D is the same  for the various examples
and near to the predicted value. The
notation for the vectors refers to $(x,y,z)$.}
\label{tdepa}
\end{table}

 If the noise power does not contain frequencies equal or higher
than those needed to go to the next set of states, the \dm should
relax  to 1/2. With  our typical parameters, this frequency is
about 0.4. In Fig\,\ref{rlx}  it is seen that when the maximum
noise frequency
is below this value the relaxation is indeed to 1/2. With higher
frequencies present the relaxation is to less than 1/2. (That it
falls to  1/4
reflects the fact that  the numerical calculations were carried out
in a 4-state basis.)

 Although our various noise power spectra do not have a simple
rectangular cutoff, we anticipate a similar behavior with our
exponentially cutoff planckian spectra. We do indeed find with low
T's and couplings that are not too strong that calculations
performed with a two-state basis give the same results as  basis
with more states, indicating that the excitation of higher states
is not important. The results we present below are always under
these conditions.

\subsection{State Independence of D}
 Probably the most striking aspect of the theory is that, with the
given assumptions, the effect of the environment, here represented
by the noise, can be represented by a single parameter $D$. Most
importantly,  D does not depend on the state of the \syn. This
means that \eq{pdota} is a {\it linear} equation for \bp. This
would not be the case, for example, if one had $D=D(P)$.  Linearity
for the evolution of the \dm is a general result of \qm and it is
non-trivial that our approximate method respects this.

We can provide some tests of this feature in the full numerical
simulation. One may take a given noise signal with a given coupling
and determine D for different configurations of the potentials and
of the initial state. 
In Table\,\ref{tdepa} we show some examples, using `z-noise' with
the parameter $\delta \eta =0.00128$, for which, according to
\eq{znoise}, the
prediction is D=0.00146.

 The first two entries are for symmetric
potentials, $x^{ext}=0$, that is with \bv in the x-direction. For
the
first line the \sy is started
with \bp also in the x-direction, that is with a \wf which is a
 linear combination of equal  wavepackets in the left and right
potential wells. In the second line the \sy is started with \bp in
the z-direction, corresponding to the initial \wf concentrated in
only one potential well.
In the next two entries we show the case of some asymmetric 
potentials, so that \bv lies at some angle in the x-z plane. In the
third line \bv is at a 45 degree angle, and in the fourth line
close to the z-axis at a  12  degree angle.

\begin{table}
\begin{center}
\begin{tabular}{|l|l|l|l|} 
\hline
\bv&$T$&$D$~~~& $\rho_{11}$\\
\hline
\hline
(0.0044,\,0.0,\,0.0)&0.0125&0.0011&0.50\\
\hline
&0.025&0.00128& 0.50\\
\hline
&0.05&0.00132&0.50\\
\hline
&0.10&0.00145&0.50\\
\hline
&0.20&0.00157&0.45\\
\hline
\end{tabular}
\end{center}
\caption{ Temperature dependence of the \de parameter D determined
from the numerical simulations using noise signals in the `z-
direction' with the power
spectrum \eq{np} and coupling $\delta \eta=0.00128$. According to
\eq{znoise}, D should be
temperature independent and have the value 0.00146. The final value
of the \dm
element $\rho_{11}$ should be $\tfrac{1}{2}$ if there is no
excitation of higher states above the effective two-state \syn. The
last entry for  $\rho_{11}$ shows some deviation   from this as T
approaches the principal level spacing $0.4$.}
\label{tdep}
\end{table}
\subsection{Test for `z-noise'}\label{testz}
We now examine the nature of the evolution with ``noise along the
z-axis". With  $\bf V$  along the
x-axis,  Fig\,\ref{vpn} shows this situation with \bp in some
general direction.

\begin{figure}[htbp]
\begin{minipage}{0.46\textwidth}
  \includegraphics[angle=-90,width=1.0\textwidth]{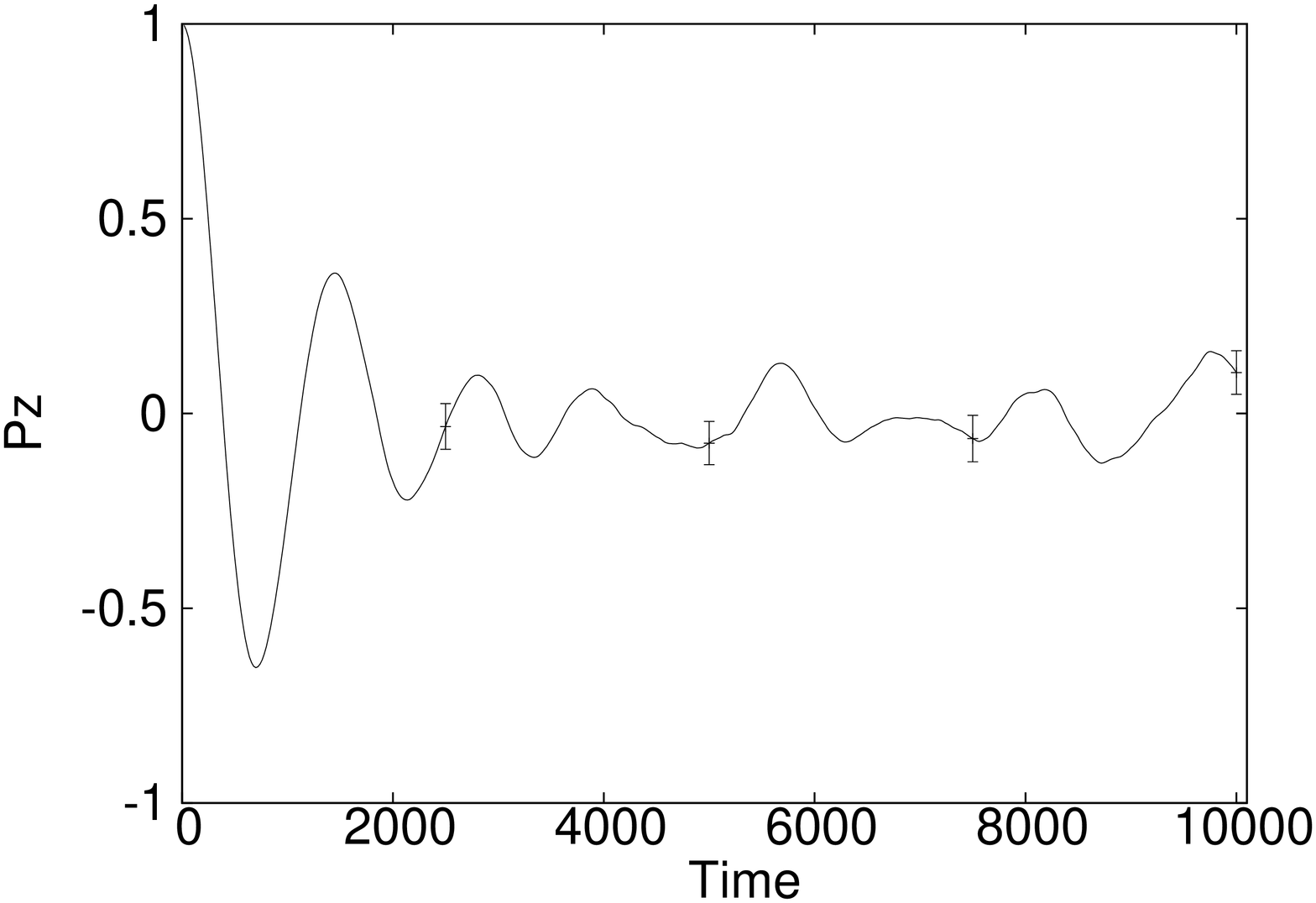}
\end{minipage}
\hfill
\begin{minipage}{0.46\textwidth}
  \includegraphics[angle=-90,width=1.0\textwidth]{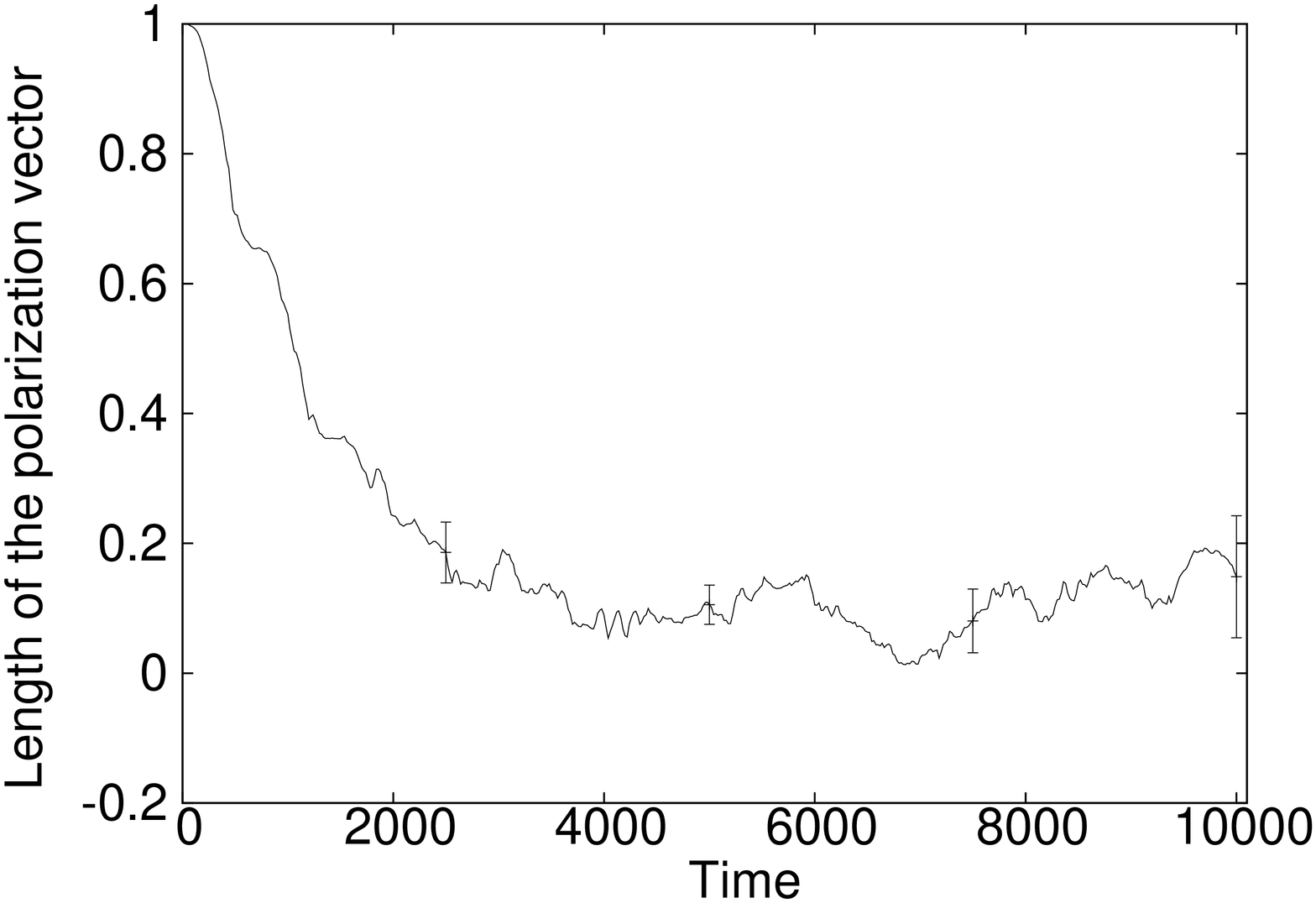}
\end{minipage}
\caption{Simulation with \eq{hama3} using ``z-noise'' due to
$\phi^{ext}$
fluctuations as given by \eq{znoi}.   The
configuration corresponds to  Fig\,\ref{vpn} with \bp started along
the z-axis.  Left:
Projection of \bp on the
z-axis showing damped
oscillations. Right: Damping of the total length of \bp. A fit 
for the length with $ e^{-Dt}$ yields $D=0.00132$, while the
prediction from
\eq{znoise} is $D=0.00146$.  The error bars  indicate
the dispersion of values expected for an average made from N
samples, namely $\frac{1}{N}\sqrt{\Sigma_1^N (P_a-{\overline
P})^2}$, where $P_a$ is the quantity being plotted, resulting from
a given noise signal `a'.} \label{znoilr}
\end{figure}

In a  first test we start \bp along the z-direction 
again with coupling $\delta \eta=0.00128$. One anticipates that
\bp will
rotate around the
x-axis while its length decreases. Inserting \eq{znoi}  
in \eq{hama3}, evaluating the \dm and extracting \bp one finds the
results  shown in Fig\,\ref{znoilr}. The left panel shows the
projection of \bp on the z-axis, with the expected damped
oscillations. The right panel shows the decrease of the total
length of \bp.

The value of D emerging from the simulation is found by fitting
$e^{-Dt}$ to the curve of the right panel, yielding  D=0.00132.
On the other hand the prediction from \eq{znoise}  is  D=0.00146.
Thus there is good agreement.

These plots were produced using the n=1 planckian noise \eq{np}.  
It is a feature of the prediction \eq{znoise} with fixed
 $\delta \eta$ that the D resulting from
the simulation should  be given by ${\cal P}(0)$  and  independent
of  T, even though a change of
$T$ implies a change in the frequencies present in the noise.

To test this,  Table \ref{tdep}  shows the values of D resulting
from various T's
used in  simulations with n=1 planckian noise. The first and last
values of the temperature are near the margins of the allowed
region $0.0044 <<T,\om_c<<0.4$. One observes approximate agreement
with the analytic prediction D=0.00146. At the highest temperature
the excitation to states above the lowest two states manifests
itself in the relaxation of the \dm element to a value less than
$\tfrac{1}{2}$.

These results are meant simply as a check on our numerical and
mathematical methods and
 should not be taken to mean that the \de is temperature
independent. To the contrary, a 
temperature dependence is rather to be expected. With the
normalizations we have
adopted, it  would be
represented
by a temperature dependence of the coupling parameter $\delta
\eta$.

\subsection{Test for `x-noise'} \label{x-n}
Fluctuations in $\beta$ correspond to changes in  the tunneling
\yyn, a $\sigma_x$ operator in our notation.
As was discussed in section\,\ref{xzcomp}, fluctuations in 
$\beta$  have  relatively small effects   compared to similar
fluctuations in  the $x^{ext}$ parameter, leading  to a much
smaller D for   $\beta$ fluctuations or `x-noise'.
 
If for the purpose of comparing the analytic and numerical
predictions, we were to increase the coupling constant $\delta
\beta_o$
in the simulations, there is  the difficulty that using a larger
$\delta \beta$ will induce substantial changes in the small
tunneling
\yyn, violating
our assumption that the fluctuations do not essentially change the
\syn. For example, we  work with $V_x=\om_{tunnel}=0.0044$,
and according to \eq{bx} the $B_x$ induced by a $\delta \beta$ is
$(-0.18)\delta \beta_o \,{\cal N}$. If we wish to have
$B_x<<0.0044$
with the typical size of noise signals $0.16$ (see sect.\ref{nm}),
then
we require $\delta \beta_o<< 0.01$, which implies D on the 
   $10^{-6}$ level, according to \eq{flba}.

\begin{figure}[h]
\begin{minipage}{0.46\textwidth}
  \includegraphics[angle=-90,width=1.0\textwidth]{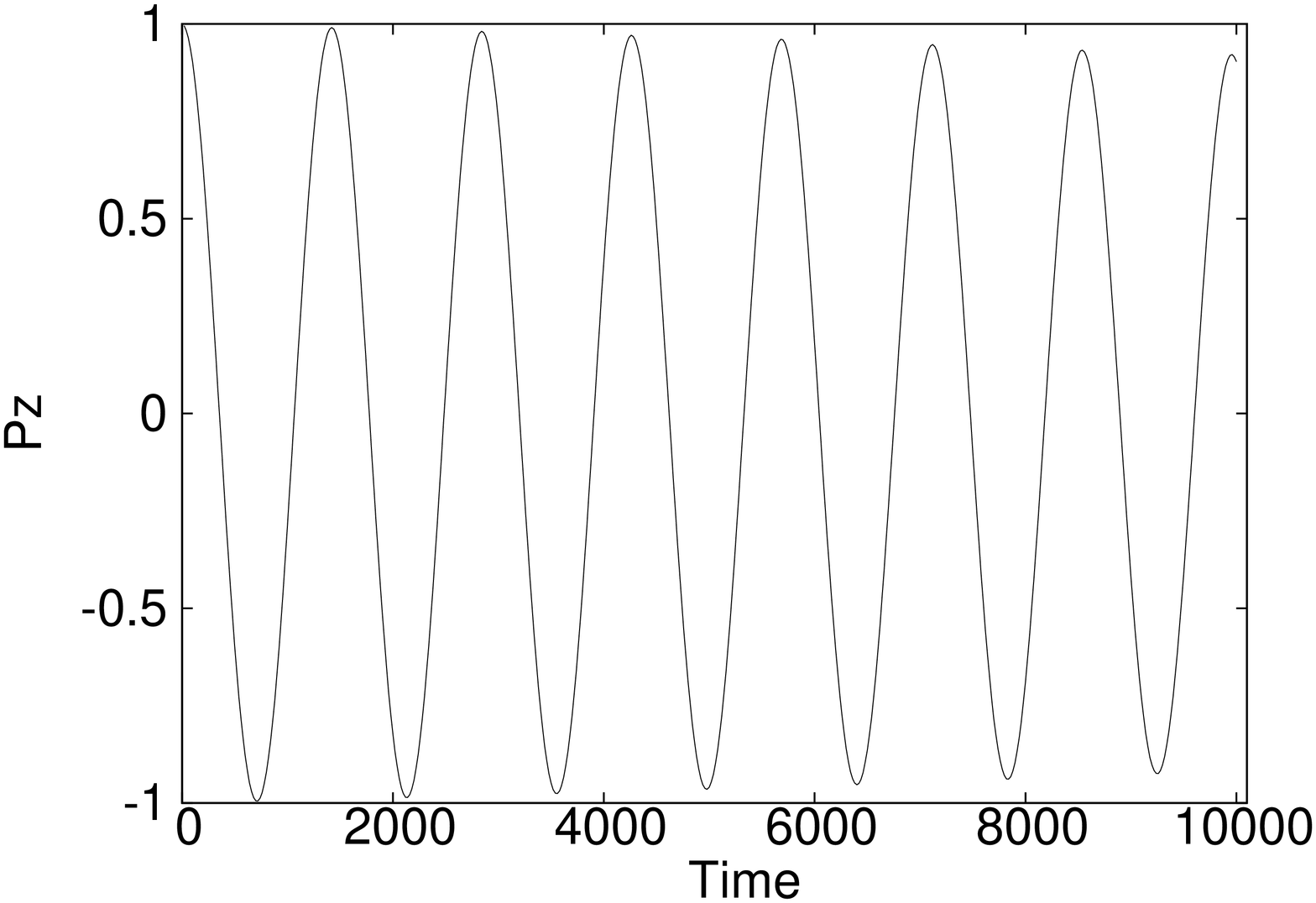}
\end{minipage}
\hfill
\begin{minipage}{0.46\textwidth}
  \includegraphics[angle=-90,width=1.0\textwidth]{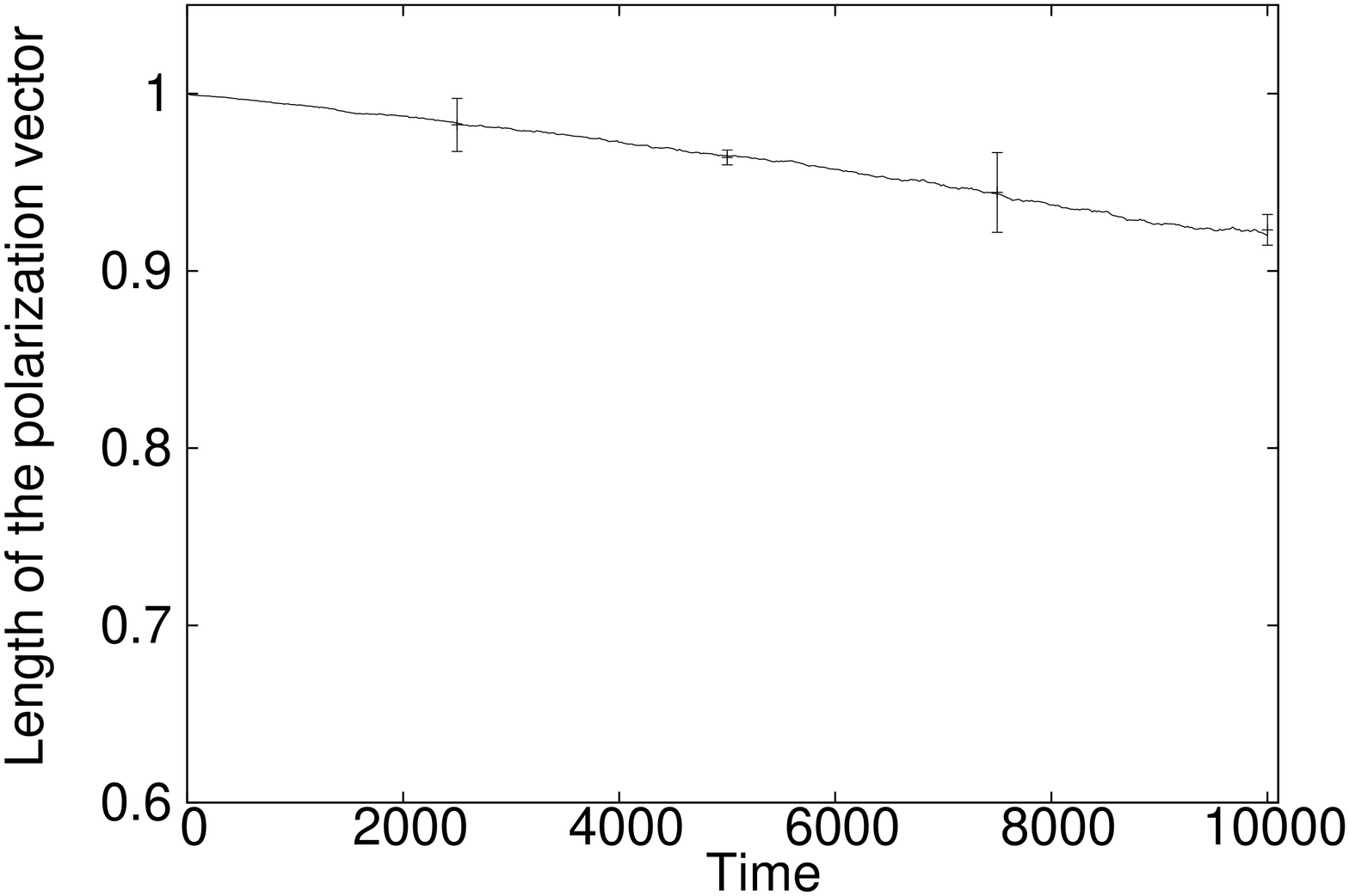}
\end{minipage}
\caption{ Relaxation due to $\beta$ fluctuations or ``x-noise''.
Left: Projection of \bp
on the z-axis. Right: Damping of the total length of \bp.
For the
configuration  of Fig\,\ref{vpn} with \bp started along the z-
axis. A fit to the length 
with $ e^{-Dt}$ yields $D=5\times 10^{-6}$, while the prediction
from \eq{znoise} is $D=3\times 10^{-6}$. A weak coupling $\delta
\beta_o=0.01$ implying weak damping must be used to avoid a large
 shift $\delta\om_{tunnel}/\om_{tunnel}$.}\label{xrlx}
\end{figure}

While such 
 small effects can be  unstable and difficult to detect
numerically, we  show in
Fig\,\ref{xrlx} a run for $x^{ext}= 0$ with the coupling $\delta
\beta_o= 0.01$. The temperature for  n=1 planckian noise was set
to
T=0.05. A fit gives $D=5\times 10^{-6}$, while the
prediction of \eq{flba} is $D=3\times 10^{-6}$. Although there
is  qualitative agreement, it is evidently  not so
precise as in the case of the z-noise. This may be due to our
various approximations as well as numerical uncertainties.

\section{``Collapse" and  ``Watched Pot''}\label{str}

Since the beginnings of \qm the \cow \quad has been assumed to be
an
instantaneous and somewhat
mysterious process, not describable in a physical manner.
However, as was noted in the beginning of  studies of \de
\cite{us}, the action
of the external influences, here simulated by the noise, may be
quantitatively examined and the  induction of a \cow-like behavior
of the \sy studied. This occurs through
the effect of the $-DP_{T}$ term,  which tends to eliminate off-
diagonal matrix elements of the \dm and so to reduce the \dm to a
classical probability distribution, an incoherent mixture of the 
states of a definite
``quality'', the eigenstates of $\sigma_z$. 

  As the external influence, represented by D, becomes very strong,
the \cow\, occurs very rapidly. In addition, the further evolution
of the \sy is inhibited and in the limit can be
essentially stopped completely \cite{us}--much like the `fixing' of
the state by a `measurement'. This is the ``Turing'' or
``Watched Pot'' effect. In terms of \eq{pdota} this occurs
because  the large $D$ ``pins''   $\bf P$
to  the noise axis so there is a
very small $P_{T}$ and  very little subsequent damping.

\subsection{Strong Damping with ${\cal N}$, \bv  Perpendicular}

With our present \sy we can  illustrate  and study such effects
quantitatively. Let the initial $\bf P$ be chosen to have length
one (pure state) and be oriented at an angle such that both $P_z$
and $P_x$ components are substantial. Fig\,\ref{vpn} indicates such
a starting
condition. If there is a strong noise along the z-axis,  $P_x$ will
rapidly disappear: there is a \cow\, to an incoherent mixture of
``up'' and ``down'' states. After this ``collapse'' the large value
of  $D_z$ inhibits  further evolution-- the ``Watched Pot''
effect.

\begin{figure}[h]
\includegraphics[angle=-90,width=\hsize]{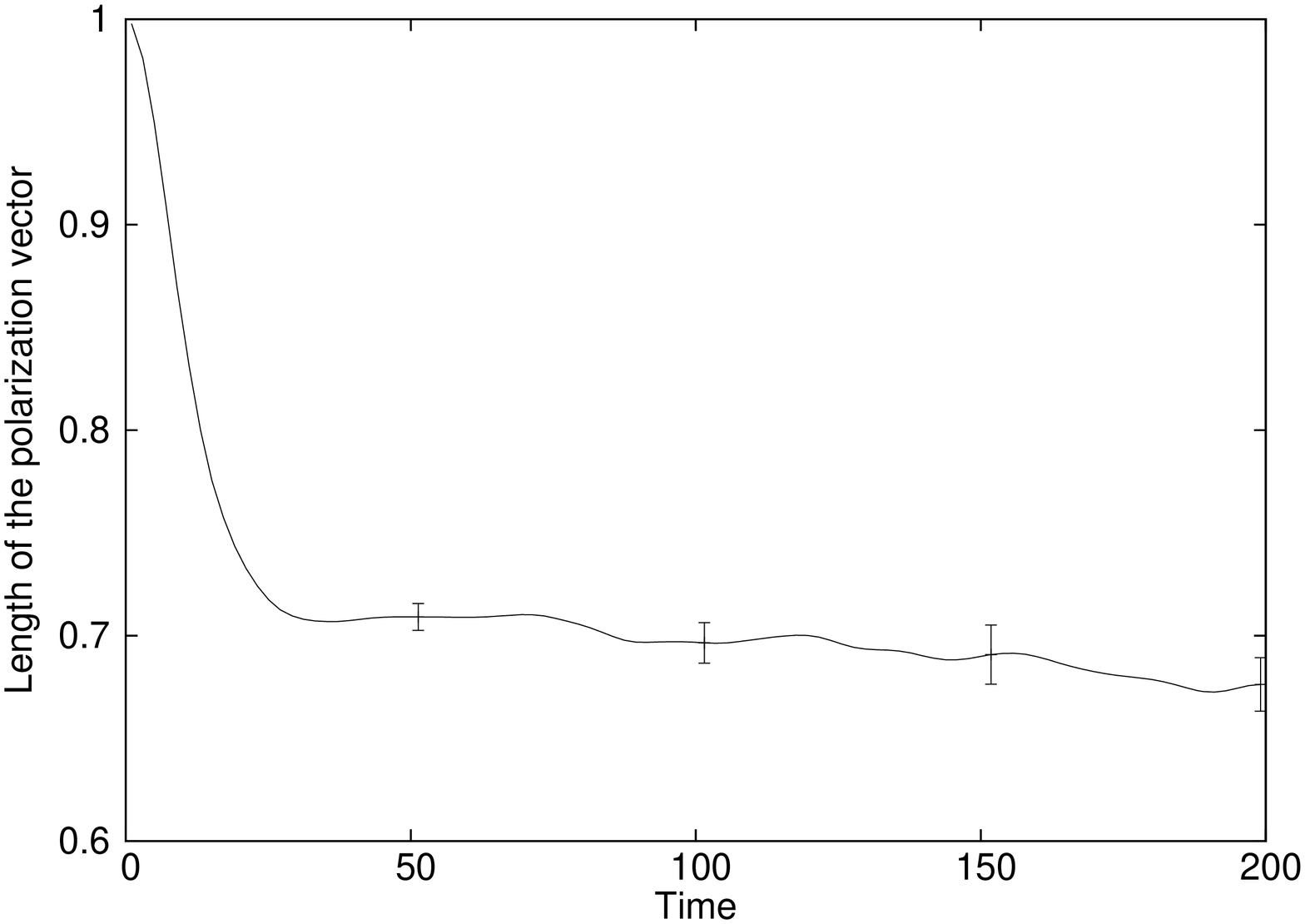}
\caption{The \cow\, followed by the ``Watched Pot'' effect. The
plot shows the evolution of the length of \bp for the configuration
of
Fig\,\ref{vpn} with the initial \bp of length one and at 45 degrees
in the x-z plane.   The $P_x$ component ``collapses''
to zero quickly, giving an incoherent mixture of ``up'' and
``down''states. This is then followed
by a near ``freezing" of the evolution. These effects result from
the
strong noise  (or ``measurement''), here with $D/\om \approx
100$. Note the short time scale, the time unit is the same as in
the other
plots, namely with  $L=400 pH$ and $C=0.1pF$ for a SQUID,  the unit
is $6.3\times 10^{-12}~sec$.}\label{wp}
\end{figure}

 Fig\,\ref{wp} shows a simulation for such conditions.
The length of \bp is plotted. One observes a rapid decrease of  the
length until only the original $P_z$ survives. The \wf 
"collapses" in a very short time . After this, the length
remains ``frozen'' to this value--the  ``Watched Pot" effect.
Increasing D gives a faster "collapse" and a stronger ``watching". 
As with the other simulations, the full \ha \eq{hama3} is used
here.

The parameter characterizing ``strong'' or ``weak'' damping is
the ratio $D/\om$, where $\om=V$ is the \yy splitting of the two
levels or the magnitude of \bv. This follows from \eq{pdota}, where
dividing through by V yields a dimensionless time parameter $t\om$
and the parameters $D/\om$.  For small $D/\om$ one has weakly
damped precession of the \bp vector around $\bf V$. For
$D/\om>>1$ one enters the ``Watched Pot'' regime.

Examination of the solutions of \eq{pdota} for \bv in the x-
direction with $V_x=\om_{tunnel}$ and \bp initially in the z-
direction leads to the behavior
\cite{us}
\beql{dlim}
P_z\approx e^{- t(V_x^2/D)}
\eeql
for \bp initially along the z-axis.
 This result follows from combining the z and y components of
\eq{pdota} to give the equation
$\ddot{P_z}+{D}\dot{P_z}+V_x^2{P_z}=0$
and solving for large D \cite{has}.
For very large D the graph in Fig\,\ref{wp} will 
decrease extremely slowly (see Fig\,6 of \cite{one}).

\subsection{Strong Damping with ${\cal N}$, \bv  Nearly
Parallel}
 In the above case $\cal N $ or $\bf B$ and $\bf V$ were at right
angles. This
is the usual ``Watched Pot''  configuration as it has been studied
for   chiral molecules,
neutrinos, or the SQUID with a
symmetric potential and dominant flux noise \cite{us} \cite{one}.
In all these cases one has-- or it is assumed--, that the
observable ``quality'' in question (chirality, flavor, flux state)
is conserved in the interaction with the environment. Since we take
this `quality' to be represented by an eigenstate of $\sigma_z$,  
 $\bf B$ is necessarily in the z-direction.

 However, one may also
envision the opposite situation where $\cal N $ and $\bf V$ are in
the same direction.
This would be the case, for example, with a strongly asymmetric
potential for the  SQUID ($\phi^{ext}\neq 0$) and dominant flux
noise. (This appears to be  the case for the experiment mentioned 
in ref\,\cite{low}.) 

 In this case the analog to \eq{dlim} for the long term
behavior of the parallel component after the perpendicular
components have been ``collapsed'' is found to be
\beql{pll}
P_{parallel}\approx P_{parallel}^oe^{-t\frac{DV_x^2}{V^2+D^2}}\,.
\eeql
 $P_{parallel}^o$ is the component remaining after the initial
``collapse'' and $V^2$ the square of \bv . This result follows
from  writing \eq{pdota} as a set of linear
equations, taking \bp $\sim e^{-\lambda t}$ and solving the
determinental
equation for the smallest eigenvalue $\lambda$.

 For $D>>V$ one recovers
\eq{dlim}. 
 For $D<<V$  there is a new regime with a  suppression due to the
large value of $V$, with the behavior $\sim e^{-t D\theta^2}$ where
$\theta=V_x/V$.
Naturally in  the limit where \bv and $\cal N$ are exactly
parallel, \bp will remain constant, as one sees setting $V_x=0$ in
\eq{pll}.

\section{Vanishing Noise Power at Low Frequency } \label{pw}
 In the numerical work so far we have assumed that the time
integral of the
autocorrelation function, or equivalently $\cal P$(0), is non-zero.
However a very interesting
special case occurs when it is in fact zero as for \eq{twth}. Then 
\eq{d} implies
\beql{vng}
D=0 ~~~~~~~~~~~~~~~~~~~~~~\int_o^{\infty}\overline{B(0)B(t)}=0 
\eeql

To check if \eq{vng}  holds  with the full \ha
we perform simulations comparing   noise signals with planckian
power spectra
for   n=1 and n=3. If \eq{vng} indeed  holds there should be
little or no
relaxation in the n=3 case.  Since our different signals have been
normalized in \eq{np} and \eq{twth} to have the same variance
$\overline{{\cal N}^2}$,
and the 
square root of the variance is the typical size of the signal, our 
 comparison is with signals of the same `strength'.

\begin{figure}[htbp]
\begin{minipage}{0.47\textwidth}
  \includegraphics[angle=-90,width=1.0\textwidth]{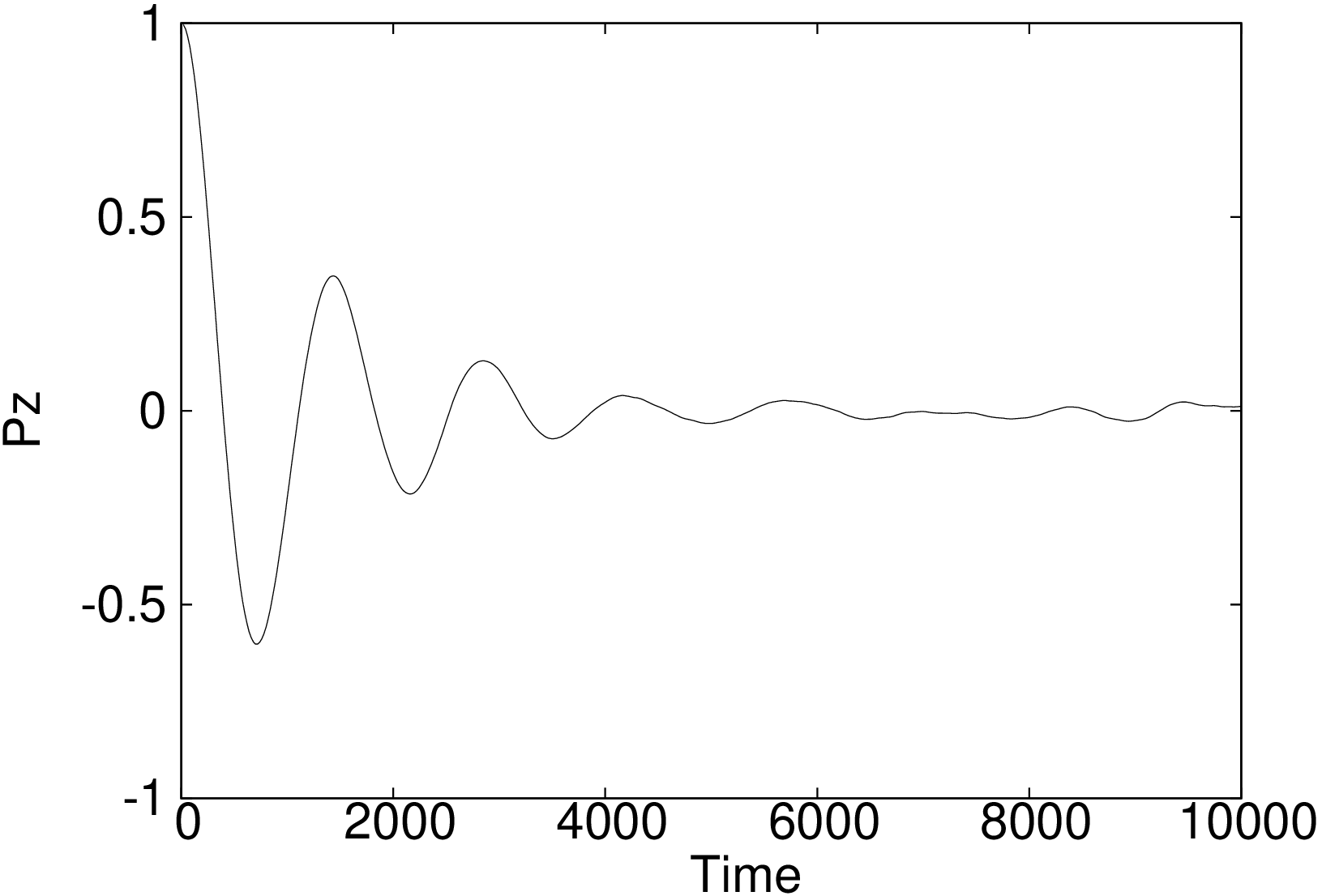}
\end{minipage}
\hfill
\begin{minipage}{0.47\textwidth}
  \includegraphics[angle=-90,width=1.0\textwidth]{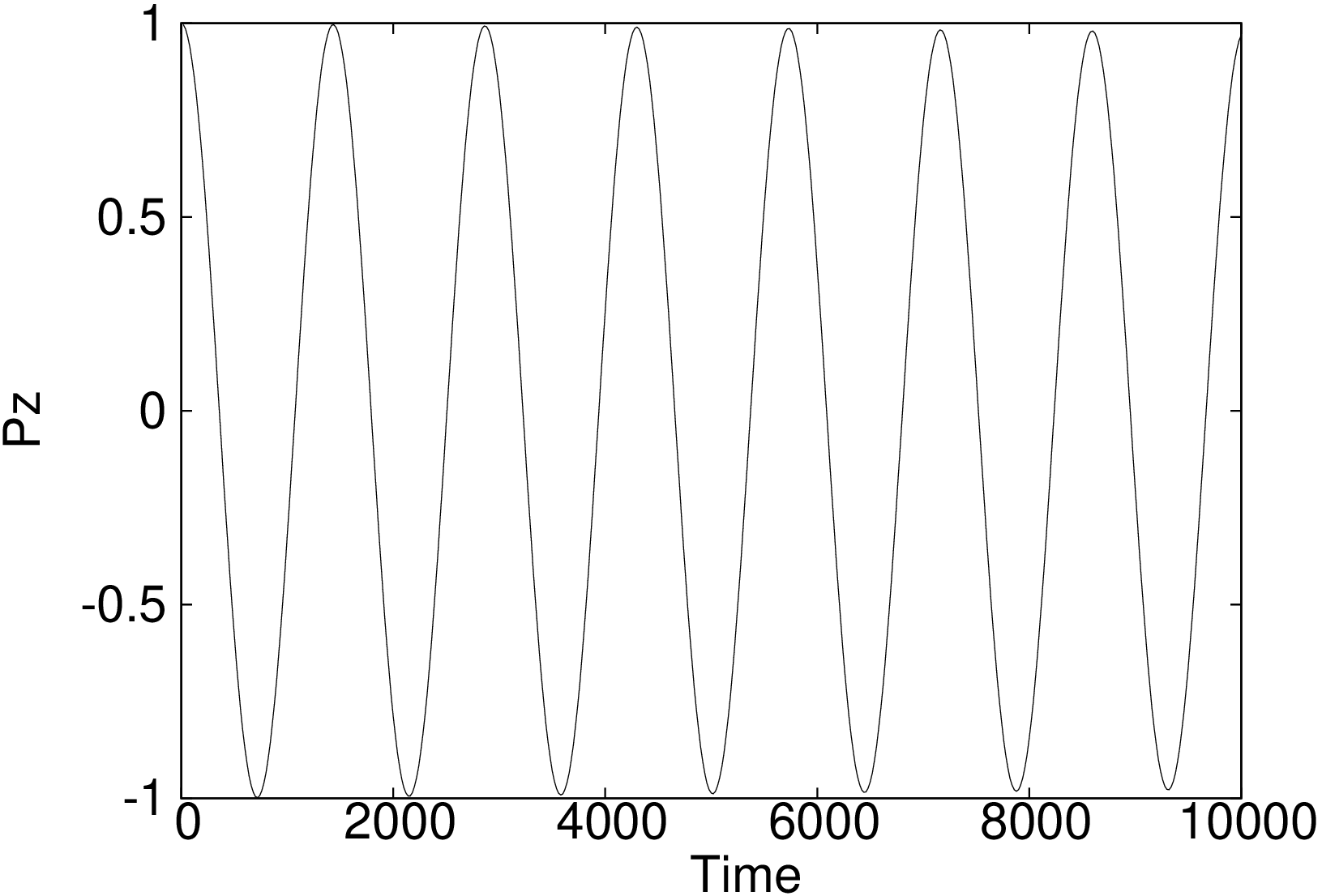}
\end{minipage}
\caption{  Identical runs  for $P_z$ but with two
kinds of noise. Left:
Planckian noise power spectrum with n=1 (lower curve of Fig\,
\ref{samp}), having
non-zero  power at zero
frequency. Right: Planckian noise power spectrum with n=3, having 
zero 
power at zero frequency (middle curve of Fig\,\ref{samp}). The two
noise
signals are adjusted so as
 to have the same variance or
integrated noise power.  
There is a striking difference in the damping or
  \den , even though the equality
of  variances assures signals of the same ``strength''. The
conditions are
the same as for Fig\,\ref{znoilr}.}
\label{twonoises}
\end{figure}

Employing these signals for noise in the z direction 
  in  Eq\,\ref{znoi}, we
obtain the results shown in
Fig\,\ref{twonoises}.  The configuration is as in Fig\,\ref{vpn}
with \bp  started in the z-direction and the parameters  the same
as for 
Fig\,\ref{znoilr}. 
The left panel  of Fig\,\ref{twonoises} is  for n=1 noise  and 
shows the expected relaxation of $P_z$.
The right panel is for n=3 and shows no apparent \de or damping
effect in the
oscillations of $P_z$.

An examination of the length of \bp shows that while it goes to
zero for
n=1, for n=3 it declines
by only  about 1\% over the length of the run.

 Thus despite the at first sight   similar appearance of the lower
two  signals in Fig\,\ref{samp} there is a great difference
in  the
\de they induce and the
expectation from \eq{vng} is indeed verified in full calculations
with \eq{hama3}.
The  exact vanishing of $D$ in \eq{vng}  is  presumably a
consequence of the
use of second order perturbation theory in the derivation 
 and should be lifted in higher order.  A similar run for n=2 shows
in fact some weak \den, with $D$ at the $\sim 10^{-5}$ level,
perhaps indicative of the higher order effects.

\section{Very Low Frequency Noise}\label{vln}
 \eq{pdota}
 gives a local-in-time description in terms of a single
``frictional" or ``dissipative'' constant $D$. $D$ does not depend
on the state of the \sy (the value of $\bf P$) or on the internal
\ha (the value of $\bf V$). $D$ depends only on the properties of
the noise, and our simulations have quantitatively verified this.
This simple structure  has its origin in the
``random
walk''  behavior of the integral of the noise signal $\int_0^t
B(t')dt'$. If $B$ is a random variable then its integral represents
a random walk and, as was  used in
arriving at \eq{defa}, one has
$\overline{(\int_0^tB(t')dt')^2}= 2t {\cal A}=2tD$, with ${\cal
A}$ the time integral of the autocorrelation function of the noise.
This  is the result of a random walk with steps given by 
$\sqrt{2 {\cal A}}$. Therefore $\int_0^t B(t')dt'$ is gaussian
distributed with a variance increasing in time as $\sim t$, which
finally leads to the exponential in time $\overline{P}=e^{-Dt}$.

If, however, the noise is dominated by very low frequencies
\cite{low} such a local-in-time description is not relevant and the
methods
presented in the above discussions should not be expected to 
apply.   For times $t$ small compared to the time
scale on which the noise $B$ varies, $\int_0^t B(t')dt'$ is not
undergoing a random walk. Indeed, in the limit a  particular
realization
$B^{a}(t)$ may be taken as effectively constant and one has 
$\int_0^t B ^{a}(t')dt'=tB^{a}(0)$ instead of the $t^{1/2}$ random
walk. The \de will now result from the average over the different 
$B^{a}(0)$.

A  simplest case to analyze in this way is that of subsection
\ref{subx} where the noise  vector $\bf B$ is parallel to \bv
(junction circuit noise for the SQUID).  \bv gives the rotation
rate of \bp in the z-y plane for a \bp started along  the z-axis. 
Thus  $V\to V+B$ with random $B$ gives a spread of the orientation
of the \bp increasing in time
and so a decreasing length of the averaged \bp or \de  according to
\beql{lag}
 \overline{P}=\overline{cos\int_0^t B(t')dt'}\approx\overline{cos
B(0)t}=\int cos Bt \, Prob(B)dB ~~~~~~~t<<t_{noise}
\eeql
where $t_{noise}$ is the time scale over which $B$ changes and
$ Prob(B)$ is the probability distribution for the $B$. With a
gaussian $ Prob(B)$ one obtains 
$\overline{P}= e^{-(1/2) \overline{B^2}t^2}$. In general  for small
times one will
expect $\overline{P}\approx 1-\hf
\overline{B^2}t^2$.

One sees that  when $t<<t_{noise}$  the  relevant
quantity  is the magnitude of the fluctuations of $\bf B$, namely 
$\overline{ B^2}$. As is evident, for example, in the discussion
around \eq{tref}, this is not the same as the ${\cal P}(\omega=0)$
entering into $D$.  Rather $\overline{ B^2}$ is given by the
integral over ${\cal P}(\omega)$, \eq{tref}.

\begin{figure}[h]
\includegraphics[angle=-90,width=\hsize]{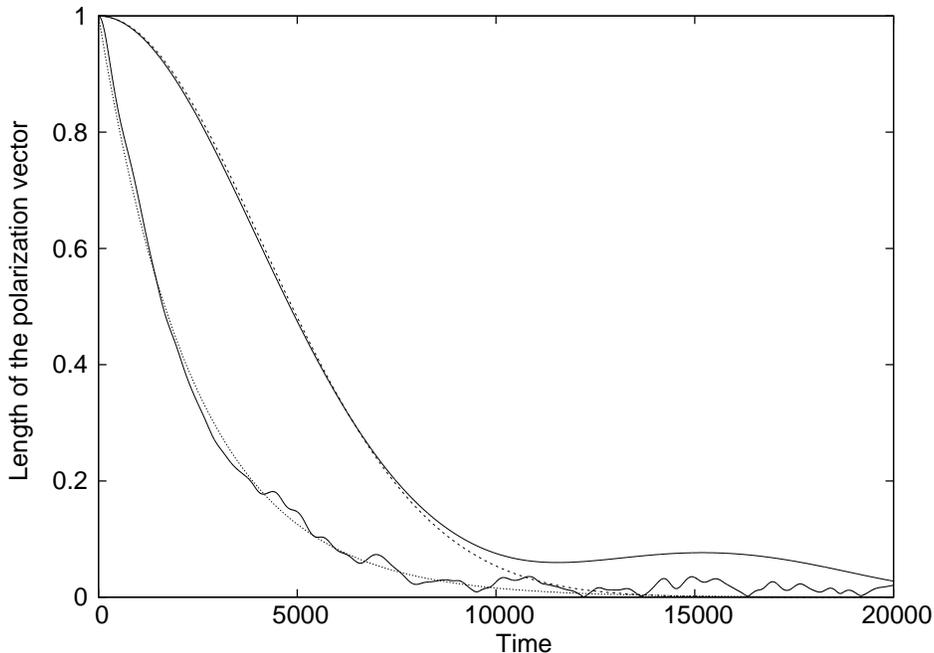}
\caption{Simulations with low and high frequency x-noise compared.
 The upper curve shows the effect of a ``rectangular'' noise power
spectrum with cutoff $\omega_c=2\times 10^{-4}$ and the lower with
$\omega_c=1\times 10^{-2}$. The low frequency case exhibits the
expected $\overline{P}\sim e^{-const. t^2}$ behavior and the high
frequency case  the $\overline{P}\sim e^{-const. t}$ behavior of
our main discussion. The dashed and dotted lines are fits for the
two types of behavior. 
  }\label{tsmall}
\end{figure}

In Fig\,\ref{tsmall} we show the result of some simulations with
very low frequency x-noise compard to a high frequency case.

 We use a cut-off white noise spectrum similar to \eq{n1}.  However
to make the situation with the noise frequencies completely clear
we use a  ``rectangular'' spectrum with ${\cal P}=1$ up to
$\omega_c$,  where it is then  abruptly cut off. The typical time
$t_{noise}$ characterizing a change in a noise signal is then less
than or on the order of $\omega_c$.  The upper  curve is for the
small $\omega_c=2\times 10^{-4}$ or $t_{noise}> 5000$. In order to
make the effects evident the relatively  large value $\delta \eta
=0.1$ was used. The behavior
$\overline{P}\sim e^{-const. t^2}$ is seen and a fit (dashed
line) gives $2.0\times 10^{-8}$ for the constant.  Using 
\eq{bx} to find the effective B, and the integral over the power
spectrum to find the variance, one predicts  $const.=(1/2)
\overline{B^2}=2.9\times 10^{-8}$, in reasonable agreement.  The
lower curve,
on the other hand, is with the higher frequency noise
$\omega_c=1\times 10^{-2}=1/100$.  One observes the
simple exponential behavior (dotted line), with the coefficient
$D=4.1\times 10^{-4}$ in
approximate agreement  with  formula \eq{znoise} for D which gives
$D=3.2\times 10^{-4}$. That the low frequency case has the smaller
effective \de in this example is due to the fact that with the
``rectangular'' noise spectrum and $\cal P$ fixed to one below
$\omega_c$, the variance   $\overline{B^2}$ decreases as $\omega_c$
decreases (\eq{tref}). While for the purposes of comparison we have
kept $\cal P$ at $\om =0$ constant for the two cases,
the behavior under experimental conditions will of course   depend 
on the  actual   amplitude of the low frequency noise.

We now turn to the perhaps more experimentally relevant situation
of subsection\,\ref{sz} , as represented by 
Fig\,\ref{vpn}. While the  
 $\bf V$ is still along the x-axis,   the noise is now along the
z-axis, (``flux noise" in the SQUID ).  \bp is again started  along
the z-axis.

The noise perturbation  along the z-axis  leads to two
differences
with the x-noise case just discussed.  One is that the shifts in
the rotation frequency are now quadratic in $B$. One has $V\to
\sqrt{V^2 +B^2}\approx V+\hf \frac{B^2}{V}$. Thus the magnitude of
the alteration of \bv is less than in the x-noise case,
but will always have the same sign, leading to an increase of the
average frequency of rotation, by
$\frac{1}{2V}\overline{B^2}$. This average change is not relevant
to the \den, but its fluctuations, characterized by
$\overline{B^4}-
(\overline{B^2})^2$, will play the role of the $B$ fluctuations
characterized by  $\overline{B^2}$
 in the x-noise case.

 The second difference has to do with the fact that since $\bf B$
is no longer parallel to \bv, the plane in which \bp rotates is
tilted somewhat with respect to the y-z plane. This will lead to
some oscillatory effects in the projections of \bp, but not to an
increasing-with-time \de arising from the spread in rotation
frequencies.

 We thus  again concentrate our attention on the spread in rotation
frequencies.
We repeat the arguments around \eq{lag}, with the role of $B(t)$
now played by $B^2(t)/2V$. 
 \eq{lag} now
becomes
\beal{laga}
 \overline{P}&=&\overline{cos\bigl(\int_0^t
\frac{B^2(t)}{2V}\bigr) dt'}\approx\overline{ cos
\bigl(\frac{B^2}{2V}t}\bigr)= \int cos \bigl(\frac{B^2}{2V}t\bigr)
\, Prob(B)dB \\
\nonumber
 &=& cos\bigr(\hf arctan(\alpha t)\bigl)\frac{1}{(1+\alpha^2
t^2)^{1/4}}\,,
\eeal
where we have taken a gaussian $Prob(B)$ and $\alpha= \overline
{B^2}/V$. For small $t$ one has $\overline{P}\approx 1-
(3/8)(\overline{B^2}/V)^2 t^2$. As before, $\overline{B^2}$ is the
relevant parameter of the noise, but now it enters as
$(\overline{B^2})^2$ since we are now concerned with the
fluctuations of  $B^2$ instead of  those of $B$.

\begin{figure}[h]
\includegraphics[angle=-90,width=\hsize]{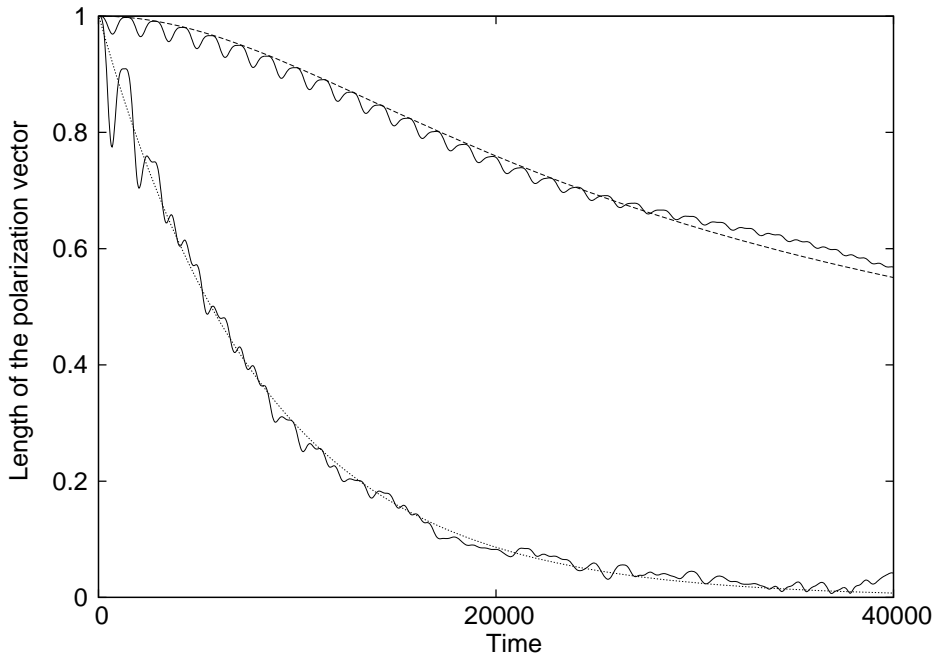}
\caption{Simulations with a very low $\om _c=8\times 10^{-5}$,  and
higher $\om _c= 8\times 10^{-4}$,  frequency z-noise.  The dashed 
line fitting the upper curve uses \eq{laga} with  
$\alpha=5.3\times
10^{-5}$. It exhibits the $1-const.t^2 $ behavior at small times,
with
$const. \sim (\overline{B^2})^2$.  The lower curve is fit with a
simple
exponential.}\label{ztsmall}
\end{figure}
In Fig\,\ref{ztsmall} we show  z-noise simulations for
$\overline{P}$ with a low, $\om _c=8\times 10^{-5}$ and a higher
$\om _c= 8\times 10^{-4}$ frequency rectangular noise spectrum. The
$\delta \eta$ parameter of \eq{bext} is $2.5\times 10^{-3}$.
One notes the above-mentioned oscillations. The  envelope to the
upper curve (dashed curve) can be fit with \eq{laga} using
$\alpha=5.3\times 10^{-5}$, while the prediction with the
parameters used is $\alpha=6.4\times 10^{-5}$. Checks on the
dependence of $\alpha$ with respect to the magnitude and frequency
of the noise signal are in agreement with the theory. The lower
curve is
fit by a simple exponential.

The fact that the \de here depends on $\overline{B^2}$ or the
integral over ${\cal P}(\om) $ leads to  amusing consequences in
the n=2,3 ``planckian noise'' cases, where the  ${\cal P}(0)$
determining $D$ is zero. As the \T is lowered and $\overline{B^2}$
becomes the relevant quantity, the  \de could increase. We find
this is indeed the case in  simulations,
with the \de increasing 
substantially  for \T $few \times 10^{-3}$, to above the 
$D\sim 10^{-4}$ level. 

A further interesting point to be noted is that  in the
general formalism the length of \bp is always entropically
decreasing (see \eq{pdotc}); while here, as in the oscillations in 
Fig\,\ref{ztsmall}, \bp can in fact increase.

In general concerning
very low frequency noise, we can conclude that   there will be
a spread of oscillation frequencies or ``dephasing''  at times
$t<<t_{noise}$, leading to a
\de which  is related to the variance of
the noise signal $\overline{B^2}$, as opposed to the integral over
the autocorrelation  function  determining $D$. It appears in
different ways
according to the type of noise, as in our examples of x- or z-
noise. While with high frequency noise the $t<<t_{noise}$ regime
may be unobservably small, with very low frequency noise
it could dominate the relaxation of P, as for the upper curve of 
Fig\,\ref{tsmall}.

 In this section we have retained the assumption of a small noise
signal: $B/V <<1$ and treated a constant V. In the low frequency
limit for the noise it would also be possible to treat a large $B$,
and a slowly moving V, using the adiabatic approximation, but this
is beyond the scope of the present discussion.

\section{Conclusions }

In summary, we have reviewed and extended the formulas relating
random noise signals and the \de parameter for the effective two
level system formed by
 the lowest levels of a double potential well. The random
noise is applied 
 as a fluctuation of various parts of the full
 multi-level  \han.  
 Numerical simulations verify the theoretical predictions, in which
the relevant characteristic of the noise is the power at zero
frequency. In particular  different kinds of
noise signals are applied and we  find
reasonable agreement   between the simulations and the 
analytic predictions.

 This agreement includes  the size of the \de
parameter, as well as its independence on  the state of the \sy and
on the frequency composition of the noise. Effects related to the
excitation of levels beyond those constituting the two qubit-like
states are discussed.  The results confirm the possibility of
describing the dynamics with an equation with a
single \de constant D.  \cow\, behavior as well as that of  the
``Watched Pot Effect" are induced for large values of D. In general
one finds that, within the allowable parameter ranges, the full
complexity of the SU2 structure involving  the noise vectors, the
\bv and
the \bp arises from the two-state reduction of the full \han. A
particularly striking result, in agreement with the theory,
 is the small size or absence of 
\de when the noise signal has no power at zero frequency.

 The noise power at zero frequency appears in
fundamental treatments \cite{fk} \cite{leg} of the dissipation
problem as the level of white noise.
 As is familiar from Johnson
noise, for example, the  existence of a classical dissipative
parameter like
resistance  is associated with  a constant noise
spectrum at low frequency. We thus arrive at the
perhaps obvious interpretation that to have a small $D$ such
classical dissipative effects or parameters should be kept to a
minimum.

An interesting  general question concerning our approach
 is to what extent classical random noise
 can be used as an adequate 
simulation of \den. 
 A real, classical
signal, that is  an 
unquantized \cite{fy} field, causes as many transitions `up' as
`down' between any pair of states. This means that
 ``spontaneous
emission'' or  ``relaxation'',
where,  even in the absence of any noise, an upper  level
 emits a quantum (e.g. a phonon,
or in vacuum a photon)
and transits to the lower state is absent. On the other hand, the
neglect of 
 spontaneous emission may  not be important in practice. Quantum
logic devices should have short time scales of operation, while 
spontaneous emission rates will be slow. Study of this question 
would require a detailed analysis for specific devices.

While our main results are  in the context of standard
 high frequency noise, we also
offer a short discussion and simulations of noise signals dominated
by low frequencies. It is found and verified by the  simulations
that in
this case the relevant parameter becomes the variance or integral
over the power spectrum of the noise.

%\nopagebreak

%\section{Conclusions}

\section{Acknowledgements}
We would like to acknowledge the participation of A. G\"orlich in
the development of  the software set.

This work was partially  supported by the International PhD
Projects Programme  of the Foundation for Polish Science within
  the European Regional Development Fund of the European
  Union, agreement no. MPD/2009/6.


\begin{thebibliography}{00}


\bibitem{one}  V. Corato, P. Silvestrini, A. G\"orlich, P. Korcyl,
L. Stodolsky, and J.Wosiek, Phys. Rev. {\bf B75} 184507 (2007);
arXiv:cond-mat/0611445. 

\bibitem{units} For more on the physical units for the SQUID
and their translation into other terms  see the section  ``Squid
Hamiltonian'' of  \cite{one}.


\bibitem{jac} J. Wosiek, {\it Nucl.
Phys.} {\bf B644} 85 (2002); hep-th/0203116.


\bibitem{cmplt} See the discussion  ``Hilbert space completeness" 
in \cite{one}. 

\bibitem{bnorm} In order that the $\bf B$ appear as 
 small changes in the \bv, we have included a factor of $\hf$ in
the
definition of the random field \ha which was not present in
\cite{one}. This different normalization of the $\bf B$ accounts
for the
factor 4 difference  between \eq{d}  here and the equivalent Eq 23
of \cite{one}.

\bibitem{bfact} Alternatively, one may use $\lambda$$\bf B$ instead
of $\bf B$ to keep track of the orders in  $\bf B$ and then set
$\lambda=1$ at the end.

\bibitem{onea} See Eq 16 of  \cite{one}.



\bibitem{oof} L. Stodolsky and P. Silvestrini,
Physics Letters {\bf A280} 17-22  (2001),
arxiv:cond-mat/0004472.

\bibitem{kh} I. B. Khriplovich and V. V. Sokolov,
 Physica {\bf A}, 73, (1987).

\bibitem{chand}See for example the relation  connecting the
diffusion constant and the velocity autocorrelation function on
page 251 of {\it Introduction to Modern Statistical Mechanics}, by
David Chandler, Oxford University Press, 1987.

\bibitem{has} R. A. Harris and  R. Silbey, J. Chem Phys. {\bf 78}
7330 (1983).


\bibitem{note} One might be tempted to 
try to evaluate  the  $\phi^2$ factor in \eq{add} by using our
association $\phi\approx \phi_c \sigma_z $. This would evidently be
wrong, since it leads to no change in the \yy splittings. The
reason for
this is that the evaluation of the matrix elements of the $\phi^2$
operator goes beyond the two-state model, and involves the whole
Hilbert space of the $\phi$ variable. Our problem provides a nice
example of how in an effective theory --here the two-state \ha
\eq{tst} --
one obtains a simplification at the price of some unknown
parameters, (e.g. matrix elements of $\phi^2$) which
are only determined in the more complete theory--here the full \ha
\eq{hama3}.


\bibitem{low} For recent discussions  of dominant 
low frequency noise for SQUIDs see M. H. S. Amin and D. V. Averin,
Phys. Rev.
Lett. {\bf 100} 197001 (2008) and R. Harris, M. W. Johnson, et al. 
 Phys. Rev.
Lett. {\bf 101} 117003 (2008). 
It should be noted that the transition rate formulas in these
papers are those of the ``Turing'' or ``Watched Pot " effect. That
is, the rate is given by the tunneling \yy squared
divided by the noise/\de parameter, as in our \eq{dlim}.  As
explained in Ref\,\cite{us}
this arises as simply the solution of our Eq\,\ref{pdota} for
large D:  $P_z\approx 1-
(\om^2_{tunnel}/D)t$. Since $\om_{tunnel}$ in Harris, Johnson,
et al. appears to be quite small the stabilization of the state
could be very strong. ( For simulations see Fig 6 of 
Ref\,\cite{one}. ) Then sub-dominant effects, not necessarily
related to the tunneling interaction, could induce transitions. 
Thus
an alternative explanation of the data  of  Harris, Johnson et al.
 might be
possible where the ``Watched Pot Effect'' keeps
the \sy in  the upper state long enough  that a slow 
``spontaneous emission'' to the ground state  can  be observed.
 A test of this interpretation could be carried out by performing
the procedure of Ref\,\cite{oof}, namely  a
determination of  $D$ by means of  an adiabatic sweep of
$\phi^{ext}$ through the ``resonance'' or ``level crossing''.



\bibitem{us} 
  R.A. Harris and
L. Stodolsky,   {\it Phys. Lett.} {\bf B116} 464 (1982).
 For a general introduction and review of these concepts see
L.~Stodolsky, ``Quantum
Damping
and Its Paradoxes"
in {\it Quantum Coherence}, J. S. Anandan ed.
World Scientific, Singapore (1990).



\bibitem{ent} \me ``Coherence and the Clock'', section X, in {\it
Time and Matter} I. Bigi and M. Faessler, eds., World Scientific,
(2006); quant-ph/0303024 .  The positivity of the $DP_T$ terms
corresponds to the positivity of a certain tensor in the general
case, see  ref\,\cite{kh}. 




\bibitem{fk} G. W. Ford and M. Kac, Jnl. Stat. Phys. {\bf 46} 803
(1987). We note that in view of our observation of the importance
of the spatial dimension through its effect on the noise power
at $\om=0$, the assumption of working in one dimension in this
reference may  not be entirely innocent.

\bibitem{leg} A. O. Caldeira and 
A. J. Leggett, Physica {\bf 121 A} 587 (1983).

\bibitem{fy} See the discussion in R.P. Feynman and A. R. Hibbs,
{\it Quantum Mechanics and Path Integrals}, McGraw-Hill (1965),
around Eq\,12-113
concerning  the reality of the $\alpha$ parameter in the
influence functional formalism.


\end{thebibliography}
\end{document}